\pdfoutput=1
\documentclass[usenatbib,fleqn]{mn2e}
 \usepackage{ifthen} \newboolean{pubformat} \setboolean{pubformat}{true}
\usepackage{natbib}		% bibliography
\usepackage[draft]{pdfpages}

\newcommand{\continuefloat}
{\ifthenelse{\boolean{pubformat}}{\contcaption{}}{\ContinuedFloat}}
\ifthenelse{\boolean{pubformat}}
{
\usepackage{fixltx2e}     %preserve figure order when 1/2-column figures mixed
\newcommand{\onecol}[1]{\includegraphics[width=84mm]{#1}}
\newcommand{\twocol}[1]{\includegraphics[width=168mm]{#1}}
}{
\usepackage{times}
\newcommand{\sun}{\odot}
\newcommand{\onecol}[1]{\includegraphics[width=6.5in]{#1}}
\newcommand{\twocol}[1]{\includegraphics[width=6.5in]{#1}}
}

\usepackage{graphicx}
\usepackage{subfig}

\title[Generation of mock tidal streams]
{Generation of mock tidal streams}
\author[M. A. Fardal et al.]   %mnras
{Mark A. Fardal$^1$\thanks{E-mail: fardal@astro.umass.edu}, 
Martin D. Weinberg$^1$, 
Shuiyao Huang$^1$\\
$^1$Dept.\ of Astronomy, University of Massachusetts, 
    Amherst, MA, 01003, USA
}  %don't put a \\ after last author
\date{Submitted to MNRAS \today}
\pagerange{\pageref{firstpage}--\pageref{lastpage}} \pubyear{2014} %mnras

\ifthenelse{\boolean{pubformat}}
{
\voffset -1.4cm    %for astro-ph
\setlength{\textheight}{241mm}  
}{
\setlength{\textwidth}{6.5in}
\setlength{\oddsidemargin}{0in}
\setlength{\evensidemargin}{0in}
\setlength{\textheight}{227.6mm}  %9in - 1 mm
\topmargin=0mm
\headheight=0mm
\headsep=0mm
\footskip=13mm
}

\newcommand{\Myr}{\,\mbox{Myr}}
\newcommand{\Gyr}{\,\mbox{Gyr}}

\newcommand{\kpc}{\,\mbox{kpc}}

\newcommand{\kms}{\,\mbox{km}\,\mbox{s}^{-1}}

\newcommand{\msun}{\,M_{\sun}}
\ifthenelse{\boolean{pubformat}}
{
\newcommand{\vecOmega}{\mathbf\Omega}
\newcommand{\vectheta}{\btheta}
}{
\newcommand{\vecOmega}{\mathbf{\Omega}}
\newcommand{\vectheta}{\mathbf{\theta}}
}
\newcommand{\nbody}{$N$-body}
\newcommand{\tsim}{\sim\!}
\newcommand{\rtidal}{r_\mathit{tidal}}
\newcommand{\figdir}{figures}
\newcommand{\racc}{R_\mathit{acc}}
\begin{document}
\maketitle    %mnras
\label{firstpage}

\begin{abstract}
In this paper we discuss a method for the generation of mock tidal
streams. Using an ensemble of simulations in an isochrone potential
where the actions and frequencies are known, we derive an empirical
recipe for the evolving satellite mass and the corresponding mass loss rate, 
and the ejection conditions of the stream
material.  The resulting stream can then be quickly generated either
with direct orbital integration, or by using the action-angle
formalism.  The model naturally produces streaky features within the
stream. These are formed due to the radial oscillation of the
progenitor and the bursts of stars emitted near pericenter, rather
than clumping at particular oscillation phases as sometimes
suggested.  When detectable, these streaky features are a reliable 
diagnostic for the stream's direction of motion and encode other 
information on the  progenitor and its orbit. We show several tests
of the recipe in alternate potentials, including a case with
a chaotic progenitor orbit which displays a marked effect on the 
width of the stream.  Although the specific
ejection recipe may need adjusting when elements such as the orbit 
or satellite density profile are changed significantly, our examples 
suggest that model tidal streams can be quickly and accurately
generated by models of this general type for use in Bayesian sampling.
\end{abstract}

\section{INTRODUCTION}
Tidal streams are interesting as potentially very high-precision
probes of the mass and shapes of galactic potentials. They also mark
the deaths of galactic satellites (which extends the sample of dwarf
galaxies we can study), serve as a link between intact galaxies and
the smoother halo component, test the predicted population of dark subhalos
within galaxy halos, and provide insight into the formation of
globular cluster systems.

The methods used for modeling tidal streams
range in difficulty from simple orbit fits to use of \nbody\ models.
The accuracy of the analysis method should bear some
relation to the quality of the observational data and the visible complexity 
of the stream under discussion.
Some tidal streams appear simply as a slightly broadened track, 
while others appear to have multiple components 
(as in the case of the Sagittarius stream, \citealp{belokurov06} and
\citealp{sohn14}) 
or density variations in excess of random noise
(as in the case of Pal 5, \citet{odenkirchen03}.
Obtaining a fit to a given stream can be useful,
but it is more informative to obtain confidence intervals for
parameters of interest, which usually involves Bayesian sampling
techniques that require many stream-model comparisons.

Some recent papers have proposed stream analysis methods that avoid
specific modeling of the tidal disruption process
\citep{pricewhelan14,sanderson14}. These methods work well 
when given high precision information on all six
dimensions of phase space. But in most cases, and especially when some
dimensions are unconstrained by observations, it seems useful to have
the analysis method incorporate more information on the way the stream
stars are released from their progenitor. This will be even more
advantageous in cases where substructure from radial oscillations is
detectable in the stream.

Among those methods that actually model tidal streams for comparison
to data, \nbody\ simulations are the most accurate way to treat the
problem, but also the most expensive, as well as in some ways
the most difficult to interpret.  Therefore their application has
been limited to date \citep{howley08,fardal13}. Simpler methods
generate a single track to be fitted to the stream. Many authors have
simply assumed that the stream follows the orbit of the progenitor,
which is a useful approximation but not a generally correct one. Other
methods for generating a stream track include the streakline
techniques of \citet{varghese11} and \citet{kuepper12}, which release
particles at constant intervals. These methods are most useful when
the progenitor is slowly and continuously disrupting. Other, more
approximate methods for estimating the stream track which are useful
in different circumstances have been used as well
\citep{johnston98,fardal06}.

Action-angle methods, as used by numerous authors
\citep{helmi99,eyre11,sanders13a, sanders13b, bovy14, sanders14}
result in the clearest conceptual picture of the dynamical structure of the
stream. However, these methods are equivalent in their function to
doing orbit integrations in the host potential, and therefore still
require accurate release conditions for the particles.  Some 
recipes have been proposed within this context \citep{eyre11,bovy14},
but they have not yet been shown to be precise or generally applicable. 
A separate issue is that computation of 
action-angle coordinates requires separable potentials for 
maximum efficiency, and if that
is not possible actions and angles may not exist globally for the
given potential.

Recently, \citet{gibbons14} and \citet{bonaca14} have both proposed
modeling the streams using sprays of particles released from the
satellite and integrated within the host potential. \footnote{Very
recently, N.~Amorisco posted a preprint covering some of the 
issues discussed here (astro-ph/1410.0360).}
\citet{gibbons14}
assert the potential of the satellite must be included to make this
feasible. Using somewhat different release conditions,
\citet{bonaca14} find it unnecessary to include the satellite
potential. However, the testing presented for their method is quite
limited and one may question its general applicability. Both these
prescriptions also assume a constant progenitor mass, despite the fact
that stream creation implies an ongoing loss of mass.

In this paper, we test a recipe for stream particle ejection against a
sample of \nbody\ simulations. We judge our recipe for release
conditions by whether it reproduces distributions of actions and
frequencies computed from the simulations, 
since these are the quantities most relevant to the
observed structure of the stream. To make this comparison practical,
we conduct all the test simulations in an isochrone potential, since
the actions and frequencies in this potential are analytic functions
of the positions and velocities. To keep the testing simple, our
progenitor satellites are simple one-component models such as might be
expected for globular clusters.  The recipe can be
generalized to more complex cases with minor changes.

In Section~\ref{sec.behavior}, we review the basic properties of tidal streams,
and define the parameters controlling the way stream stars are released.
We describe the sample of simulations in Section~\ref{sec.sims}. 
In Section~\ref{sec.model}, we constrain the mass loss recipe 
which determines how many stream stars are
created and at what times, 
as well as the release parameters,
which determine the orbits of the stars once created.
Section~\ref{sec.mockstreams} illustrates the performance of the generated streams.
In Section~\ref{sec.discussion}, we discuss the wider applicability of our 
results and compare to some other recent 
stream modeling, and we summarize our results in Section~\ref{sec.conclusions}.

\section{THE BEHAVIOR OF TIDAL STREAMS}
\label{sec.behavior}
\subsection{Stream generation in phase space}
\label{sec.phasespace}
Although descriptions in other coordinate bases are often possible,
the most universal way of describing a tidal stream is as a
collection of tracer particles thrown off a satellite as it orbits its host
potential, with those particles integrated in position-velocity space
(phase space) under the combined influence of the host and the satellite.
In the case of a circular orbit within a spherical potential, 
a necessary condition for particles to escape
from a satellite at radius $r$ is that they exceed the tidal radius (or Jacobi
radius), which can be written as
\begin{equation}
\rtidal = \left( \frac{m_{sat}}{f \, M(<r)} \right)^{1/3} r
\label{eqn.rtidal}
\end{equation}
where $M$ is the mass of the host in which the progenitor orbits, and 
\begin{equation}
f \equiv 1 - \Omega^{-2} d^2 \Phi / dr^2 = 3 - d\ln(M)/d \ln(r)
\label{eqn.ftidal}
\end{equation}
and $\Omega = V_c(r)/r = (G M(<r) / r^3)^{1/2}$ is the circular angular speed
\citep{bt2}.  
Note that $f=3$ for a Kepler potential and $f=2$ for a logarithmic halo. 
We assume $m_{sat}(<\rtidal) \approx m_{sat}$, which
is commonly satisfied at least after tides have had a chance
to strip the outermost particles. Even in this case the
escape process is not simple, and much attention has been devoted to
the orbits of stars in the classic point-mass Hill problem and its
generalizations \citep{heggie01hill}. Stars escape into leading and
trailing streams from near the Lagrange points located at $\pm
\rtidal$ from the satellite along the radial vector to the host
center. 

For an eccentric orbit, there is no known rigorous definition of a
tidal radius. but the concept is still useful in
estimating the amount of mass loss. We continue to define the tidal
radius by equation~\ref{eqn.rtidal}, where $\Omega = V_c(r)/r$
is the circular angular speed.  Thus the generalized tidal radius
depends only on the potential, satellite mass, and radius. Note that some authors
cited here instead set $\Omega$ in equation~\ref{eqn.ftidal} to be the actual
angular velocity $v_t / r$ of the satellite, where $v_t$ is the
tangential velocity.

If the stripping is not too violent, the stars again 
escape into well-separated leading and trailing streams from near the
Lagrange points. A greater eccentricity of the orbit leads to a
disruption rate more strongly peaked near pericenter. If the tidal
forces are strong enough, the entire satellite may disrupt at once,
giving a continuum of ejected particles from leading to trailing.

\subsection{Stream behavior in action-angle space}
\label{sec.actionangle}
The {\em simplest} description of tidal streams occurs when the stream
stars can
be described by action-angle variables \citep{helmi99, eyre11,bovy14}. 
In these coordinates the stars behave as free particles, at
least if we ignore the continued effect of the satellite as in the
previous subsection.  There are certainly cases where an
action-angle description is {\em not} possible. For example, the stars
may become unbound from the host potential, and hence cease to execute
oscillatory behavior.  When the potential is irregular, as is true
for almost any realistic galaxy potential, actions do not truly exist.
However, the action-angle description continues to be useful, if only 
as an approximation.  For example, in a spherical halo whose potential
is flattened by the disk, an action-angle computation is likely to be effective
as long as the orbital plane precession frequency remains much smaller than 
either of the radial or azimuthal frequencies.  

Here we summarize the action-angle approach, mostly adopting 
the notation of \citet{bovy14}.  We take the current time at which the
stream is observed to be $t=0$.
Stars are released from the progenitor at 
positive lookback times $t_s$,  (times $t=-t_s$).
The Hamiltonian is expressed in terms of the actions and is 
independent of the angles.
Therefore the actions are constant with time, and the gradient of the 
Hamiltonian in action space gives the time derivative of the angles,
$\dot{\vectheta} = \partial H / \partial \mathbf{J} = \vecOmega(t_s) = \mbox{constant} $,
implying 
$\vectheta = \vectheta(t_s) + \vecOmega \, t_s$. 
 With reference to the progenitor's quantities $\vectheta^{(p)} $, 
$\vecOmega^{(p)}$, the offsets are measured as 
$\Delta \vectheta = \Delta \vectheta( -t_s) + \Delta \vecOmega \, t_s$, 
$\Delta \vecOmega = \mbox{constant}$.
Normally, we can assume that $\Delta \vecOmega$ is small (comparable
to $r_t / r$, so that
a linear expansion of the frequencies is applicable.
This will be adequate unless the progenitor mass 
exceeds say $10^{-3}$ of its host mass \citep{bovy14}, 
in which case nonlinear terms can become important.
Normally, $\Delta \vectheta(t_s)$ is small as well, so that the properties of 
$\Delta \vectheta$ are dominated by the second term.
However, in rare cases, the orbit may be close enough to circular
so that the emitted stars are ``trapped'' near pericenter and apocenter,
rather than following the phase of the progenitor.
We will return to this case later.

The orbital behavior of the stars must be 
coupled with a description of the rate at which particles are released
into the stream, and their initial action and angle offsets.  
Some authors 
focus on models with a continuous release of particles
from an initial time to the present (we call this the ``leak case'').  
If the mean frequency of
particles released into the stream is $\vecOmega^m$, then this forms a stream
track governed by 
$\Delta \vectheta \approx \Delta \vecOmega^m t_s$
where $t_s$ takes on a range of values.
Some authors instead focus on streams formed instantaneously at a 
single disruption time $t_d$ (``burst case''), in which case the track 
is instead described by
$\Delta \vectheta \approx t_d \Delta \vecOmega$
where here $\Delta \vecOmega$ takes on a range of values.
The formalism of \citet{bovy14} has a smooth disruption starting at an finite
lookback time $t_d$, so the stream track derived there smoothly 
interpolates between these two limits at a scale
$\Delta \vectheta \sim \Delta \vecOmega^m t_d$, resembling the leak case
at smaller angles and the burst case at larger ones.

Unless the orbit is nearly circular, we expect the stars to be
released predominantly but not entirely near pericenter.  If the
progenitor manages to survive its first pericenter intact, the release
model should actually resemble a combination of the leak case with a
series of overlaid bursts.  The ejection from some satellites might have 
finite starting and ending times, though in this case we expect the
scale of the frequency offsets to change throughout the lifetime of
the progenitor.

Although the distribution in action-action space shows a complex
bow-tie shape \citep{eyre11}, 
the distribution in frequency-frequency space is nearly diagonal 
(see \citealp{bovy14}).
The distance along the diagonal is correlated generally speaking with 
the magnitude of the action offset
(or more specifically, nearly with the particle energy; \citealp{johnston98}).
This leads to two qualitatively different types of ``tilts'' between
the stream and the progenitor orbit.  
First, for particles released in a single burst, the action and
frequency offset are strongly correlated.  Particles with high
energy lag behind in the orbit, while particles with low energy
speed ahead.  This creates a tilt between the stream and the orbit.

The second kind of tilt results from the fact that the narrow diagonal
distribution of particles in frequency space is in most cases not
pointed exactly along the frequency vector of the satellite
\citep{eyre11}, though in general it is at least close. One factor in
this misalignment angle is that for typical galactic potentials the
ratio of radial to azimuthal frequencies tends to decrease for stars
executing larger-radius, higher-energy orbits. Therefore, regardless
of whether
particles are released in a burst or leak slowly, the angle vectors
tends to extend along a line misaligned from the frequency vector.

The action-angle and angle-angle tilts are generally both in 
operation, so that it is not always simple to guess
where the stream will lie in real space relative to the orbit
without detailed calculations.
However, in young streams (less than a few orbits),
we can usually expect the action-angle tilt to dominate
the offset between the stream and the orbit. The opposite is true
for old streams, say those that have executed tens of orbits.
Naturally, an ejection recipe that correctly reproduces the 
distribution in release time and action space should correctly
reproduce the effect of both of these tilts.

\subsection{Substructure in the stream}
\label{sec.substructure}
Orbits of galactic satellites and globular clusters are typically
highly eccentric.  Ratios of apocenter $r_a$ to pericenter $r_p$ 
of about 4 are typical, 
as expected for mildly radial tracer populations in typical 
halo potentials \citep{vdbosch99}.  Since tidal forces are
a major influence on the mass loss rate, and these forces depend on
the radius to a high power ($r^{-3}$ for Kepler or $r^{-2}$ for a 
logarithmic halo), we can expect 
the mass loss to proceed as a series of bursts, at least for the
progenitors that do not fall apart in the first encounter with their host.
Thus most tidal streams can be expected to fall somewhere between the
extremes of ``burst'' and ``leak'' behavior.  This is one possible source
of substructure within the stream.  

\citet{kuepper08} and \citet{just09} drew
attention to the substructure in tidal streams. These models used
progenitors on circular orbits, but subsequent papers 
\citep{kuepper10,kuepper12} involved
eccentric orbits. A point possibly obscured in these papers is that
the origin of substructure in these two cases is fundamentally
different. In the circular case, the substructure arises from clumps
of stars at particular values of their {\em current} radial angle,
i.e. the pericentric locations of the stream stars.
This is only possible because the circular case is in the ``trapped''
regime mentioned above: the phase of the emitted stars in the trailing (leading)
tail is always near pericenter (apocenter), rather than continuously
increasing following the phase of the progenitor.  In the more general case,
the substructure is characterized by particular values of the 
radial angle at {\em emission}, not its current value.   
This behavior is clearest in action-angle space, where the stars
move in their tori with constant angular velocity.  
In real space, the streaky features (as do all parts of the stream)
have their density enhanced at apocenter, due to the slower motion
of the stream particles there.  The density seen in the phase-space
stream is a combination of these two effects.

The dividing line between the
trapped and untrapped regimes occurs when the 
stars emitted at apocenter are placed on circular orbits.
The conditions for this can be found using the epicyclic approximation.
We write the azimuthal frequency as
$\Omega$ and the radial frequency as $\kappa$.  
Let us suppose that stars are ejected at a position 
offset by $k_r \rtidal$ from the satellite, 
with a tangential velocity offset by $k_{v\phi} \Omega \rtidal$ from the satellite.
To place the ejected stars on circular orbits, the 
radial epicycle amplitude $X$ of the satellite must take the value
$X = (2 \Omega^2/\kappa^2) [k_{v\phi} - (1 + d\ln\Omega/d\ln R) k_r] \rtidal$,
neglecting second-order terms.  For a flat rotation curve this 
simplifies to $X = k_{v\phi} \, \rtidal$.  In other words, the epicyclic radius
dividing trapped from untrapped regimes is of order $\rtidal$, with the
exact value depending on the values of $k_r$ and $k_{v\phi}$ 
(which will be investigated in section~\ref{sec.ejectparams} below)
and the rotation curve.  This means that only nearly circular orbits 
(with $r_a / r_p \approx 1+2X$) can be in the trapped regime.
For globular clusters with a mass $\tsim 10^{-6}$ that of their host, this demands
an epicyclic radius $\tsim 10^{-2}$ that of the orbital radius, or a radial
action $J_r < 10^{-4} L_z$.   For representative dwarf galaxies, this
criterion rises to $J_r < 10^{-2} L_z$.
Thus the trapped regime probes a negligibly small region of phase space.

In the more common untrapped case which occurs for larger eccentricity,  
the clumps seen in simulations spaced fairly evenly in phase along the stream 
cannot be due to their current radial phase.  
Instead, they are determined by the radial phase at ejection,
and have two simultaneous causes: the variation of the mass loss rate with
radial phase, and the correlation of action and frequency values with radial phase.
The following example illustrates these stream features.

\begin{figure}
\onecol{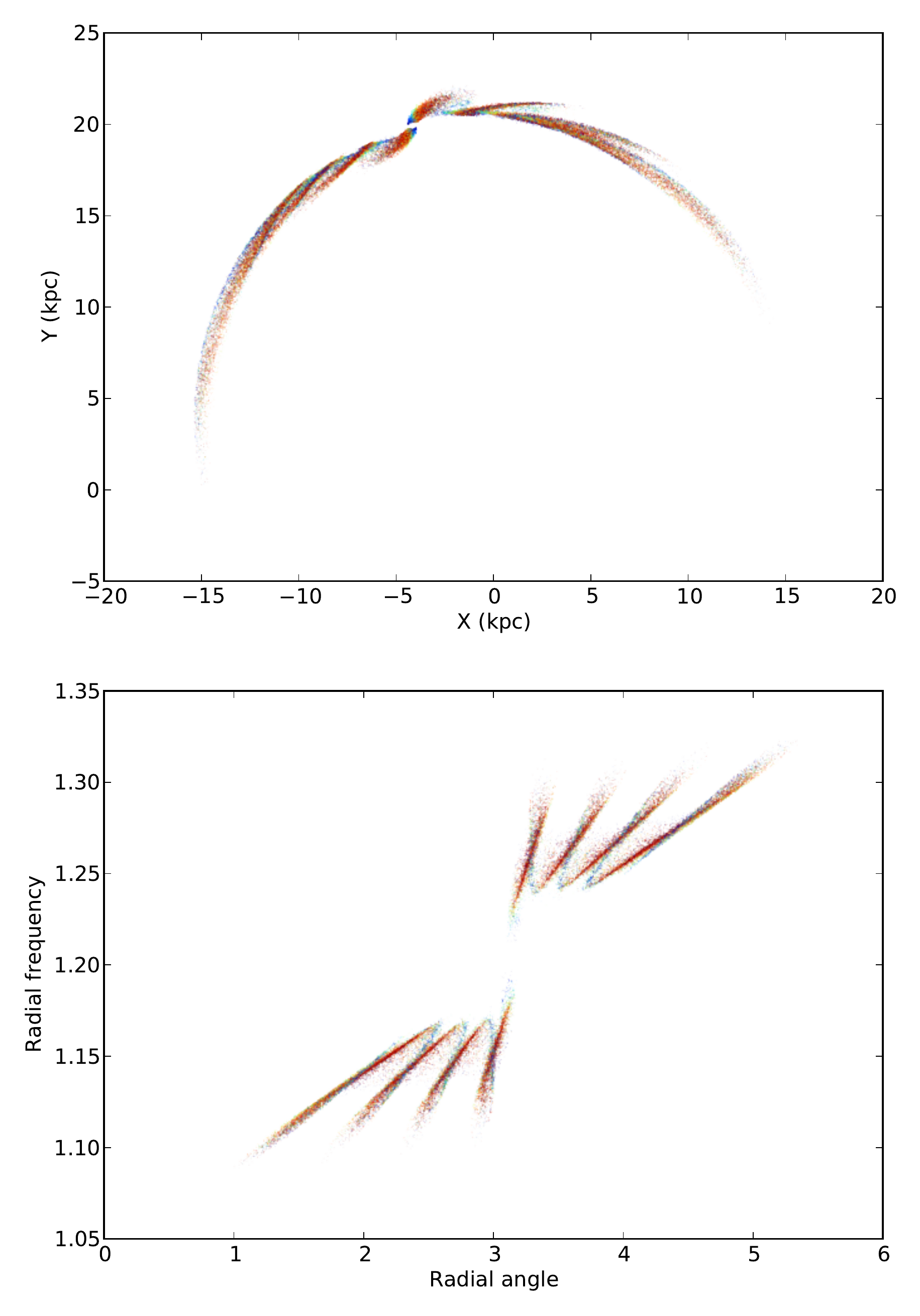}
\caption{
\label{fig.streaks}
Top: particle positions in the $X$-$Y$ plane in the {\tt orb1.5\_m6.0\_fat} run,
described in section~\ref{sec.sims} and Table~\ref{table.sims}.
Only unbound particles are shown, at the fourth apocenter
passage of the satellite.  Color indicates the cosine of the 
estimated release phase
of the satellite, with red marking pericenter and blue apocenter.
Bottom: radial frequency versus radial angle, shown at the same time as
the upper panel.
}
\end{figure}

Figure~\ref{fig.streaks} (upper panel) shows the particle locations
for one of the simulations described in Section~\ref{sec.sims}. This
simulation is shown at the fourth apocenter passage. We can see
streaky features within the stream, which increasingly overlap as we
go outwards from the satellite. If we convert to action-angle space
(lower panel), the origin of the streaks is clearer. We can see that
the streaks correspond to bursts of particles released near
pericenter, as indicated by the color coding. Even though one might
expect apocentric clumps of particles, due to the long time that the
progenitor spends near apocenter, these clumps do not appear to exist
simply because few particles are released around apocenter. Clearly,
one must take both the variation in release rate and the dispersion in
the release conditions into account to model the streaks.

In action-angle space, the inclination of the streaks has a consistent
sign across both the leading and trailing streams. 
The tilt of the streaky features in action-angle space is 
$\Delta \Omega = t_s^{-1} \Delta \theta$, thus
going to zero for large lookback times. This tilt pattern
translates fairly well to the real-space view as well. 
In general streaky features are inclined such
that their leading part points inward and their trailing part outwards
when compared to the mean stream path. Thus streaky features provide
an indicator of the stream's direction of motion, which may be useful
in cases where it is not instantly apparent from other characteristics such
as kinematic measurements or leading-trailing stream asymmetries.

Counting the streaky features provides a simple indication
of the number of pericentric passages the progenitor has undergone,
and thus the timescale over which the stream stars have been disrupted.
The spacing between the midpoint of the streaky features in
angle space is $\Delta \Omega^m(r_\mathit{peri}) T_\mathit{radial}$, 
constant in time,
though the visibility decreases as time goes on due to the increasing
alignment and overlap of the features. In real space, this spacing will
fluctuate as the stream moves from apocenter to pericenter and 
its angular speed varies.
The stream tail at $|\Delta \theta| > |\Delta \Omega^m| t_d$ 
discussed by \citet{bovy14} is in essence the oldest streaky
feature in the stream, caused by the burst of particles released at the
initial pericenter.  

\subsection{Parameterization of the ejection model}
\label{sec.ejectparams}
It would be useful to be able to model the ejection from
the satellite without performing a full \nbody\ simulation.  To model
the location of the stream correctly, we need 
an accurate recipe for ejecting particles from the satellite.
This recipe must supply both the ejection rate from the satellite
and the properties of the ejected particles at the time of ejection.
Simple models for the ejection rate could
assume a single burst, multiple bursts, or a continuous leak
of particles.  A more realistic model will assign a distribution of
ejection times within a radial cycle.

After escaping from the satellite the stars feel its influence
diminish rapidly, but the influence of the satellite (and
perhaps of the tidal debris) does not stop entirely at the tidal
radius for any star \citep[as examined in][]{choi07}. 
This influence is felt especially long for stars that lap
the progenitor in azimuth, or those that have orbital resonances with
it, and also in cases where the progenitor has a large mass and thus
affects even particles at large angular separations from it.
Thus it is not quite correct to examine the conditions of
particles crossing the tidal radius and conclude that they determine
the final tidal stream structure. However, for small progenitor masses
we can still specify the
behavior of stars fairly well in terms of orbits in the host potential
alone, when ejected from {\em some} location near the satellite 
with {\em some} velocity (not necessarily the actual velocity at that point). 
Let us rewrite these ejection conditions as follows, using polar coordinates
$(r,\phi,v_r,v_t)$.
\begin{eqnarray}
r &=& r_\mathit{sat} + k_r \, \rtidal   \label{eqn.r} \\ 
\phi &=& \phi_\mathit{sat} + k_\phi \, \rtidal / r \label{eqn.phi} \\
v_r &=& v_{r,\mathit{sat}} + k_{vr} \, v_{r,\mathit{sat}} \label{eqn.vr} \\
v_t &=& v_{t,\mathit{sat}} + k_{v\phi} \, V_c(r) \rtidal / r \label{eqn.vt} \\
z &=& k_z \, \rtidal / r \label{eqn.z} \\
v_z &=& k_{vz} \, V_c(r_\mathit{sat}) \rtidal / r \label{eqn.vz} \; . 
\end{eqnarray}

These forms are similar to some previous models, in particular
the streakline models of \citet{varghese11} and \citet{kuepper12},
The primary difference between those two models is the treatment of
the tangential velocity of the ejected particles.  \citet{varghese11} 
assume a physical velocity equal to that of the satellite ($k_{v\phi}
= 0$),  while \citet{kuepper12} assume an {\em angular} velocity 
equal to that of the satellite ($k_{v\phi} = 1$, for the $k_r = 1$ assumed there).

In principle the 
six parameters $k_r, k_\phi, k_{vr}, k_{v\phi}, k_z, k_{vz}$ 
could be constants, functions of time, or random samples from distributions
that depend on time; we will assume the latter. Along with other authors we will set
\begin{equation}
k_\phi = k_{vr}  = 0
\end{equation}
and test whether this gives a reasonable model for the ejecta.  
We assume the other parameters are either constants or are
described by Gaussian distributions.  This is unlikely to be accurate
in the extremes of the distribution, but in this paper we are only
aiming for an approximation that reproduces the typical dispersions.

The ejection conditions must be coupled with a description of the
ejection rate, which will be a function of the satellite's structure
and orbit. With these two ingredients, we can integrate particles in
the host potential to predict the phase-space properties of the
stream.

\begin{figure*}
\twocol{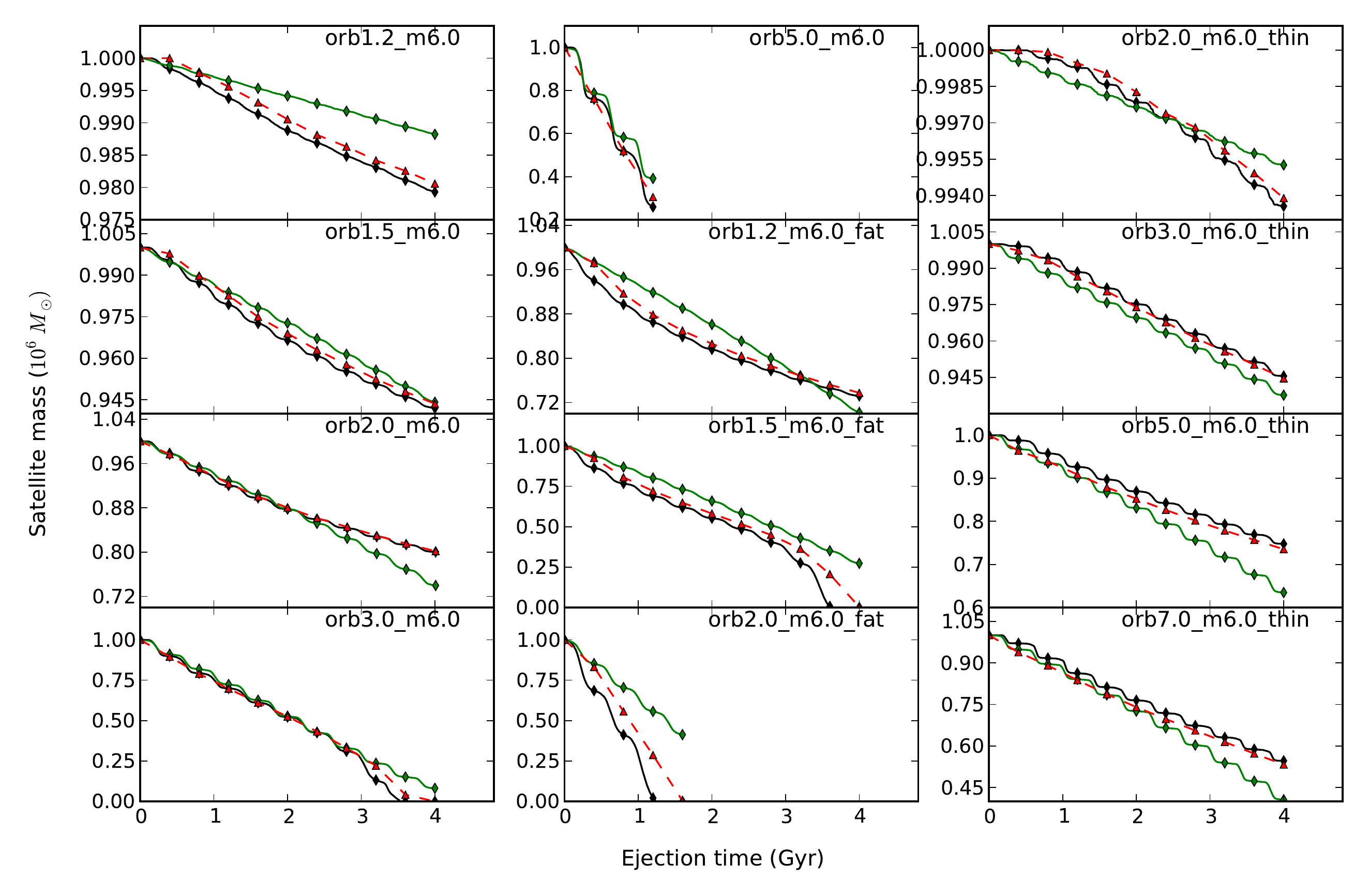}
\caption{
\label{fig.apomass}
Satellite mass as a function of time.  Black: mass 
within tidal radius measured from simulations.
Wiggles illustrating periodic mass loss enhancements
are clearly visible.
Red: predicted mass at successive apocenters, obtained
by truncating actual simulation structure at the 
previous apocenter near the pericentric tidal radius, 
as discussed in the text.  
Green: mass predicted from the full mass loss model, which
uses only the initial satellite orbit, mass and structure
as inputs.  Small symbols denote mass at apocenter.
}
\end{figure*}

\section{Simulation set}
\label{sec.sims}
To constrain our particle-spray model,
we conduct a set of \nbody\ simulations. Our goal is to
demonstrate that a reasonable model can be derived for some range of
satellites and orbits, but we do not expect that this model will be
applicable to all situations. All simulations take place within an
isochrone host potential, so we can calculate actions and angles
easily. Our simulations use relatively diffuse satellites where tidal
forces are responsible for the disruption. 
This stands in contrast to many
models of globular cluster evaporation, where ejection is 
often dominated by
internal relaxation processes and stellar mass loss; 
these must be modeled either by
collisional \nbody\ simulations or Fokker-Planck codes 
\citep{aarseth98,chernoff90,murali97,gnedin97,takahashi00,lamers10,gieles14}. 
Thus the specific fits derived here may not be applicable to all globular
clusters. However, roughly half of MW clusters (preferentially 
at small galactocentric radii) have destruction
time scales dominated by tidal forces as opposed to relaxation
\citep{heggie01massloss}.  The observable globular streams may be
predominantly those where strong tidal forces induce rapid leakage of
stars, so our models are not necessarily inapplicable to some globulars
as well as dwarf galaxies.  

\begin{table}
\caption{Test simulations}
\label{table.sims}
\begin{tabular}{lrrcrrr}
\hline
Name &$r_p$ (kpc)&$r_a$ (kpc)&$r_a/r_p$&$T_{rad}$ & $m_{sat}$ & $f_t$ \\
\hline
            {\tt orb1.2\_m6.0}&$ 16.92 $ & $  20.30$& $  1.2 $ &$400.0 $ & $ 10^{6} $ & $ 0.8 $ \\
            {\tt orb1.5\_m6.0}&$ 14.88 $ & $  22.31$& $  1.5 $ &$400.0 $ & $ 10^{6} $ & $ 0.8 $ \\
            {\tt orb2.0\_m6.0}&$ 12.40 $ & $  24.77$& $  2.0 $ &$400.8 $ & $ 10^{6} $ & $ 0.8 $ \\
            {\tt orb3.0\_m6.0}&$  9.24 $ & $  27.72$& $  3.0 $ &$400.0 $ & $ 10^{6} $ & $ 0.8 $ \\
            {\tt orb5.0\_m6.0}&$  6.12 $ & $  30.51$& $  5.0 $ &$400.2 $ & $ 10^{6} $ & $ 0.8 $ \\
    {\tt orb1.2\_m6.0\_fat}&$ 16.92 $ & $  20.30$& $  1.2 $ &$400.0 $ & $ 10^{6} $ & $ 1.2 $ \\
    {\tt orb1.5\_m6.0\_fat}&$ 14.88 $ & $  22.31$& $  1.5 $ &$400.0 $ & $ 10^{6} $ & $ 1.2 $ \\
    {\tt orb2.0\_m6.0\_fat}&$ 12.40 $ & $  24.77$& $  2.0 $ &$400.8 $ & $ 10^{6} $ & $ 1.2 $ \\
  {\tt orb2.0\_m6.0\_thin}&$ 12.40 $ & $  24.77$& $  2.0 $ &$400.8 $ & $ 10^{6} $ & $ 0.5 $ \\
  {\tt orb3.0\_m6.0\_thin}&$  9.24 $ & $  27.72$& $  3.0 $ &$400.0 $ & $ 10^{6} $ & $ 0.5 $ \\
  {\tt orb5.0\_m6.0\_thin}&$  6.12 $ & $  30.51$& $  5.0 $ &$400.2 $ & $ 10^{6} $ & $ 0.5 $ \\
  {\tt orb7.0\_m6.0\_thin}&$  4.54 $ & $  31.77$& $  7.0 $ &$400.0 $ & $ 10^{6} $ & $ 0.5 $ \\
  % {\tt sag_nfw}                   & $  $ & $  31.77 $ & $  7.0 $ & $ 400.0 $ & $ 10^{6} $ & $ 0.5 $ \\
\hline
\end{tabular}
\end{table}

\begin{figure*}
\twocol{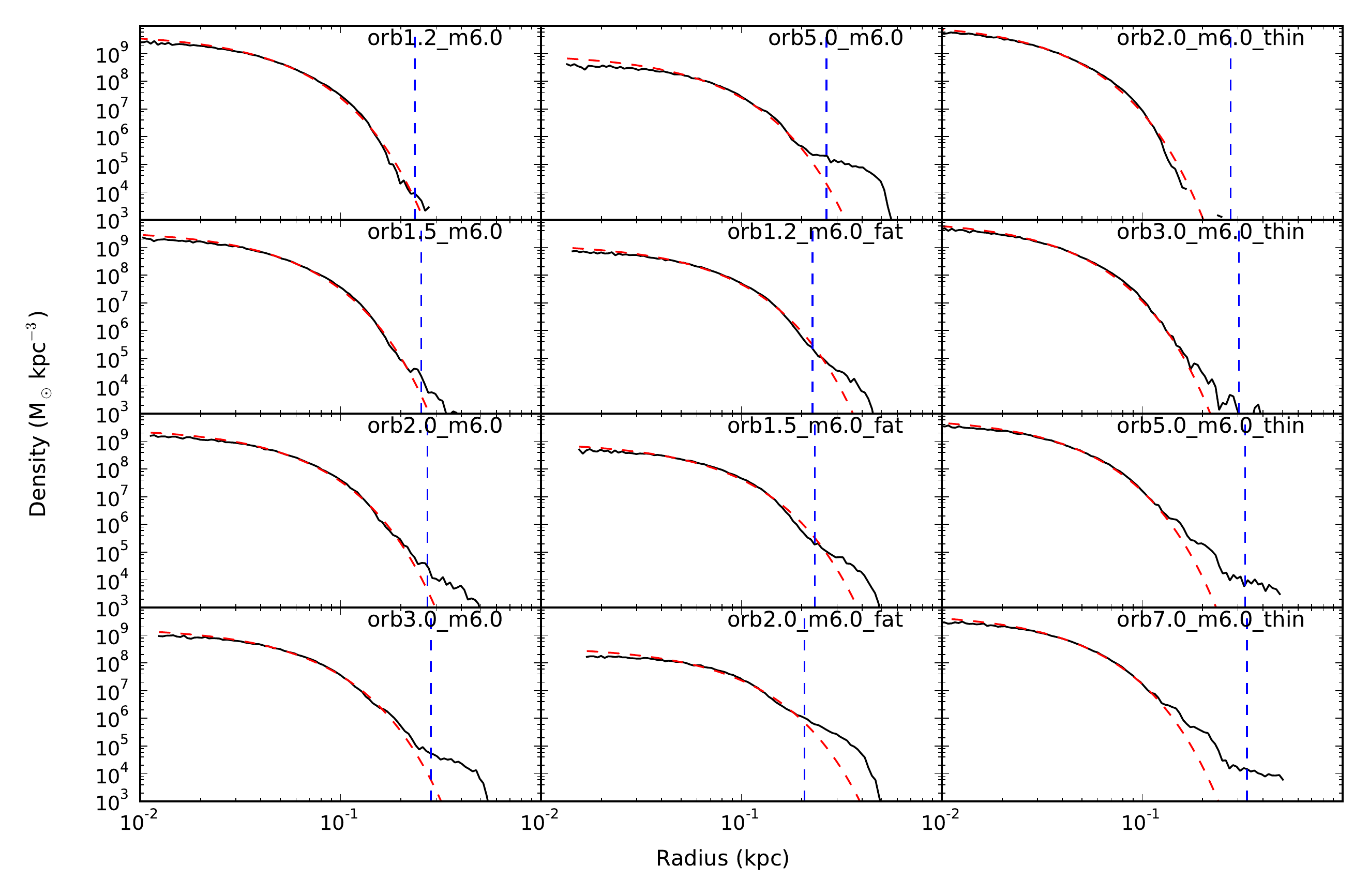}
\caption{
\label{fig.profile}
Radial density profiles, as measured at the third apocenter.  
Black: profile measured from simulation,
discarding any formally unbound particles.  
Red: profile generated from mass loss model using current mass, 
initial mass, and initial density profile.
Blue: tidal radius at apocenter.
}
\end{figure*}

Our simulations are of course scale-free, but for comparison to
observed systems we choose a unit system where the isochrone potential
has mass $M=2.852 \times 10^{11} \msun$, and scale length $b=3.64
\kpc$.   This follows the choices of \citet{eyre11} and provides a 
rotation velocity similar to that of the Milky Way for radii near that of the Sun.
All orbits are chosen to have
orbital periods 400~Gyr, and the $r_a/r_p$ span a range
1.2--7.0. We use spherical \citet{king66} model satellites with $W=3$
(i.e., not very concentrated). The simulations are listed in
Table~\ref{table.sims}, and are named according to the orbit, mass,
and scale length.
 
We place the model satellites at the apocenter of their orbit. 
We set the initial outer or ``tidal'' radius of the King model to 
be a specified fraction $f_t$ of the tidal radius at {\em apocenter}:
\begin{equation}
f_t \equiv r_{King66} / \rtidal(r_{apo}) \; .
\end{equation}
The tidal radius at pericenter can be inferred from Table~\ref{table.sims} 
and equation~\ref{eqn.rtidal}, but
in most cases it is inside the initial model outer radius,
inducing significant mass loss at each pericenter.

We run the test simulations until the time reaches 4.3 Gyr.
We save snapshots every 10 Myr.
We track the position of the satellite as a
function of time and convert it to action-angle variables. 
In some cases the orbital frequencies seem to deviate slightly from
the expectations based on the initial orbit, most likely due to 
interaction of the satellite with its own tidal debris
(as expected from \citealp{choi07}). 
Thus we compute corrected frequencies from the 
measured satellite positions, and use the new frequencies 
to initialize a finely spaced lookup table for each satellite's position 
and velocity as a function of simulation time. 

For each particle outside the tidal radius at the final snapshot, we
estimate the ejection time from its action-angle coordinates. Using
this to estimate a useful starting time, we compute the orbit of the
particle backwards in time between snapshots in a two-body
time-dependent potential, using the lookup table to compute the
satellite position.  We use this to infer a more accurate
ejection time from the satellite, as defined by the time at which the
particle crosses the tidal radius. We then save the position and
velocity of the particle and satellite. In some cases this procedure
fails. For some cases, the radial and azimuthal angles do not
predict the same ejection time. Many of these are near the satellite
and some have probably interacted multiple times with the satellite.
Some other failures occur when the satellite is falling apart rapidly
and the point mass approximation fails. When the refined approximation
fails, we find the ejection time using a linear interpolation of the
radial distance from the satellite at snapshots.

\section{DERIVING A MASS LOSS MODEL FOR TIDAL STREAMS}
\label{sec.model}
\subsection{Mass loss model}
To form a complete model of the tidal stream,
we must model the rate of mass
loss in addition to the conditions of the ejected particles.
We will do this by comparison to the simulations in
section~\ref{sec.sims}.
The first step in constructing the mass model is to specify
the satellite mass at subsequent apocenters.  Earlier work on
satellites of various types has suggested that the evolution of
the satellite mass and density profile follow quite predictable tracks,
given only the initial orbit and structure \citep{hayashi03,jorge08}.
In our simulation set, we find that the satellite mass at the next
apocenter can be reproduced quite well in most cases simply by
truncating the apocentric profile at 0.9 times the tidal radius 
at pericenter.  Figure~\ref{fig.apomass} illustrates the results
of this procedure.  The red curve represents the satellite mass 
predicted at each apocenter, based on truncating the satellite structure 
of the previous apocenter.  This agrees well with the simulation results
shown by the black curve.

\begin{figure*}
\twocol{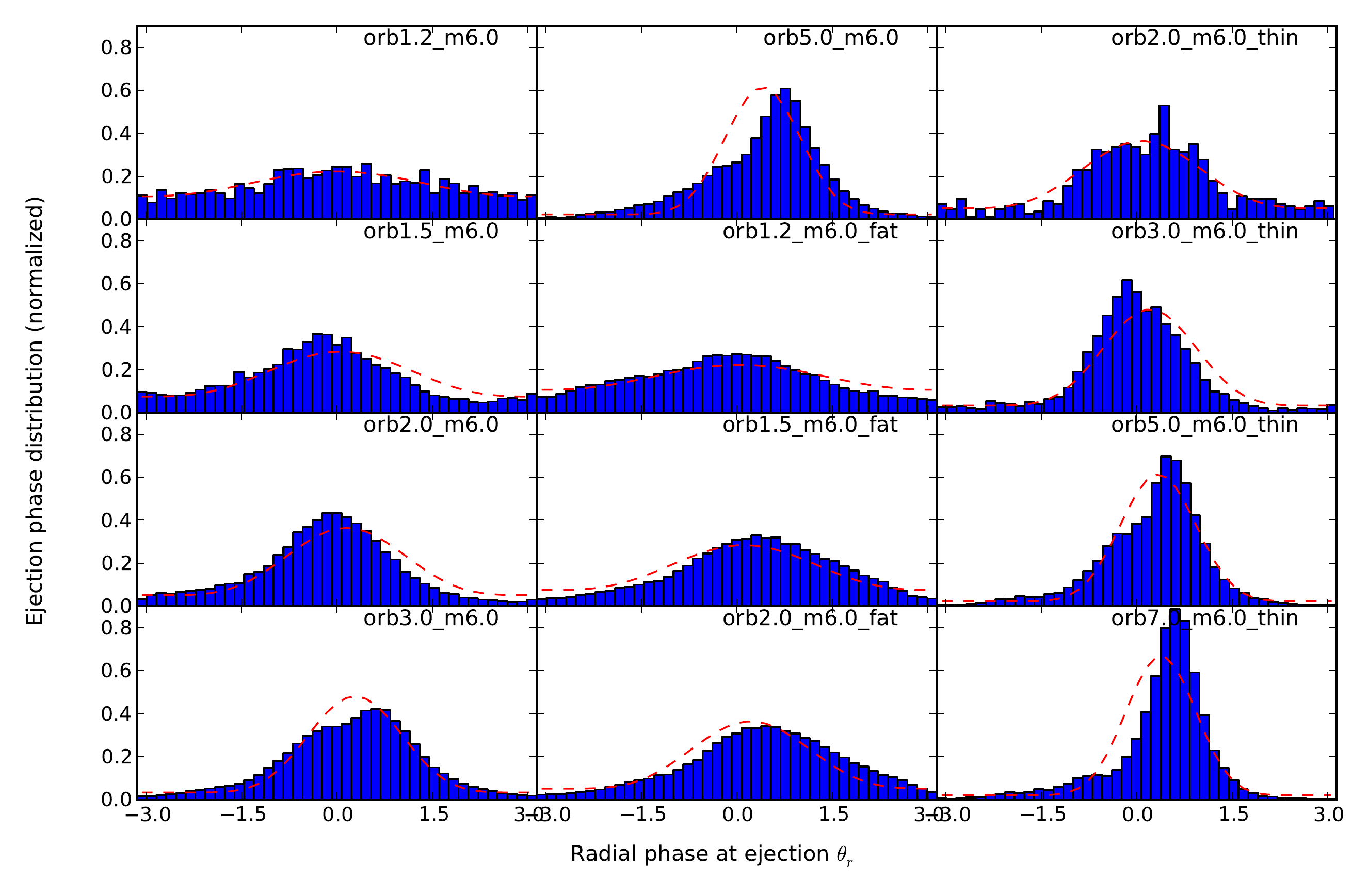}
\caption{
\label{fig.masslossphase}
Distribution of mass loss in ejection radial phase.
Blue histograms: distribution measured from simulations,
as measured from crossing at one tidal radius.
Red: model specified by orbit and initial size, as in text. 
Clearly the model captures the overall peaky behavior
while ignoring some of the finer structure.
}
\end{figure*}

However, it is not sufficient for our model to predict the 
mass using the actual, evolving structure of the simulated \nbody\ satellite, 
as we want to dispense with \nbody\ simulations
altogether.  We find that after the first pericentric passsage, the 
radial profiles of satellites that do not disrupt altogether are reasonably well 
described by Einasto profiles, as shown in 
Figure~\ref{fig.profile}.  In this figure, we only include those particles
that are formally bound to the satellite, as defined by the relative velocity 
and potential from the satellite and its debris
(and ignoring the tidal force).
We write the Einasto profile in the form
\begin{equation}
\rho(r) = \frac{(2\mu)^{3\mu} m_{sat}}{4 \pi r_{sc}^3 \mu e^{2\mu} \Gamma(3\mu)} 
  \exp\left\{ -2\mu \left[ \left( \frac{r}{r_{sc}} \right)^{1/\mu}  - 1 \right] \right\}
\end{equation}
so the index $\mu$ resembles the $n$ in the related Sersic profile.  
Here $r_{sc}$ is the radius at which the logarithmic slope of the 
density equals the isothermal slope $-2$.  For this profile, the 
cumulative mass is given by 
\begin{equation}
M(<r) = m_{sat} \; \gamma\left[ 3\mu, 2\mu (r/r_{sc})^{1/\mu} \right]
\end{equation}
where $\gamma$ is the normalized incomplete gamma function.

Furthermore, we find that the radial scale evolves little
over the course of the simulation, while the density scale
drops significantly.  Therefore we simply assume the 
radial scale is fixed at $r_{sc} = 0.2 r_{outer}$, where
$r_{outer}$ is the initial outer or ``tidal'' radius of the
King model.  
The density scale then simply scales with the initial mass.
The optimal $\mu$ varies slightly between runs and
snapshots, but for simplicity we keep it fixed at $\mu=0.9$.
In other words, the density profile is fairly close to exponential.  

The resulting fits are shown in Figure~\ref{fig.profile}.
At large radius the tidal debris shows some complicated 
structure which cannot be modeled by a simple profile, but
within a radius containing most of the mass the fit is 
fairly good.
Once we have a sequence of apocenter masses, this determines the
amount of mass lost over each radial cycle, and this prediction is shown
in Figure~\ref{fig.apomass}.

As our simulations take place in a spherical potential, we include
only the effects of ``bulge shocking'' but neglect those of ``disk shocking''.
The performance of our recipe should be checked particularly in 
cases where satellites pass directly through the disk of the host,
as these may require an enhanced degree of mass loss which is 
dependent on the full details of the orbit rather than just the
oscillations in radius.  

\begin{figure*}
\twocol{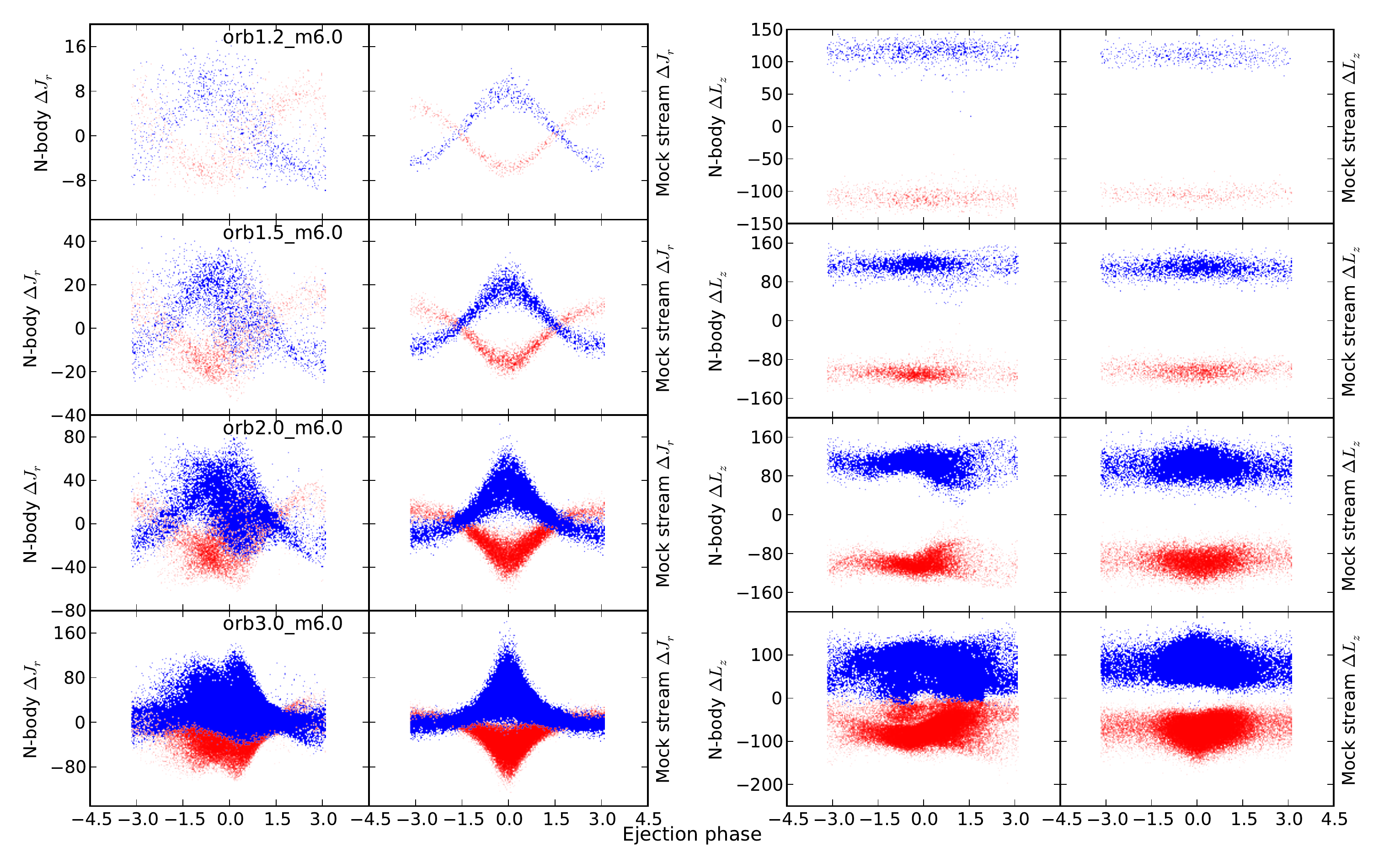}
\caption{
\label{fig.actions}
Actions of stream stars.
First two columns show $J_r$, second two show $L_z$.
Both are given in units of $\mbox{kpc}\kms$.
In each pair of columns, simulation results are on
the left and results from the mass loss model on the right.
}
\end{figure*}

To determine the variation of mass ejection rate within this cycle,
we first define the ``acceleration gradient'' as 
\begin{equation}
g_a \equiv \Omega^2 - d^2\Phi(r)/dr^2 \; ,
\end{equation}
where $\Omega = V_c(r)/r$ is again the circular frequency.  
This quantity represents the second derivative with radius
of the effective potential, for a circular orbit, and thus measures
the strength of the tidal force.  (Note $g_a = \Omega^2 f$).
We then compute the ratio of this quantity at apocenter and pericenter, 
\begin{equation}
\racc \equiv \frac{g_a(r_{peri})}{g_a(r_{apo})}
\end{equation}
We use $\racc$ as our measure for the 
variation of the tidal forces over a radial oscillation.

We use the following analytic form to define the ejection rate
as a function of radial phase $\theta_r$:
\begin{equation}
\frac{dM}{d\theta_r}=  A \left( 1 + (r_{ej}-1)
   \left[ \frac{1 + \cos(\theta_r - \theta_{mid})}{2} \right]^\alpha \right)
\label{eqn.ejectrate}
\end{equation}
With this form, the ratio of ejection at peak to trough is $r_{ej}$.
The normalization and cumulative distribution
of this form can be calculated using beta functions 
(complete and incomplete respectively).  

We then set the parameters in this expression using
\begin{equation}
r_{ej} = \exp(1.4 \, \racc^{3/4})
\end{equation}
\begin{equation}
\alpha = \racc^{0.55}
\end{equation}
\begin{equation}
\theta_{mid} = -0.1 + 0.7 \frac{f_t \racc}{7 + f_t \racc}
\end{equation}
These forms were determined by comparison with the distribution
of ejection phase in the simulations
(Figure~\ref{fig.masslossphase}).

These parameter expressions are sufficient to approximate the variation
of emission rate with satellite orbital phase, to the degree required to
specify reasonable models.  They are not well justified physically
so should be checked in cases that are far from 
Note that the rate of particle emission peaks substantially after pericenter
in some cases.  

\begin{figure*}
\twocol{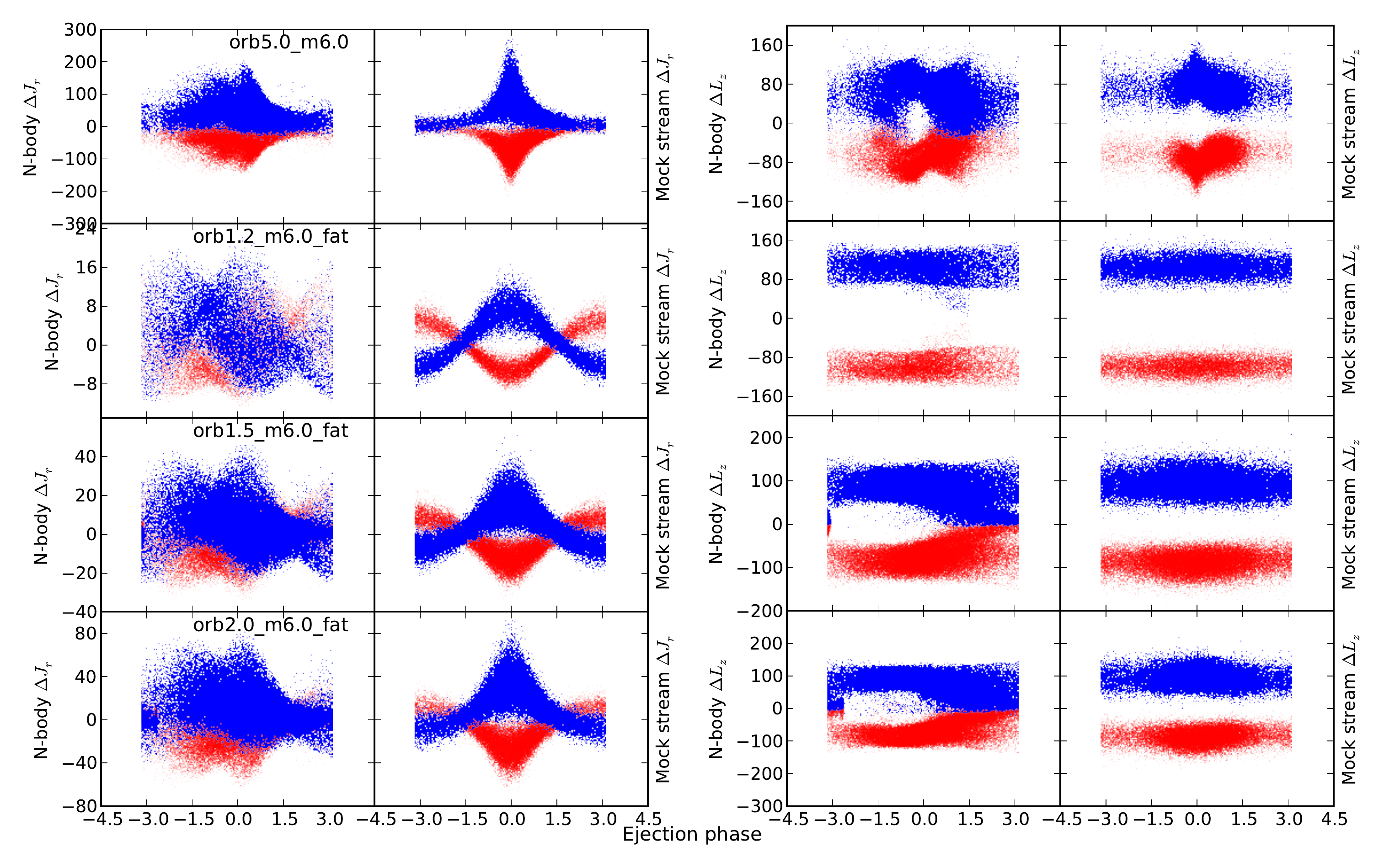}
\continuefloat
\end{figure*}

\begin{figure*}
\twocol{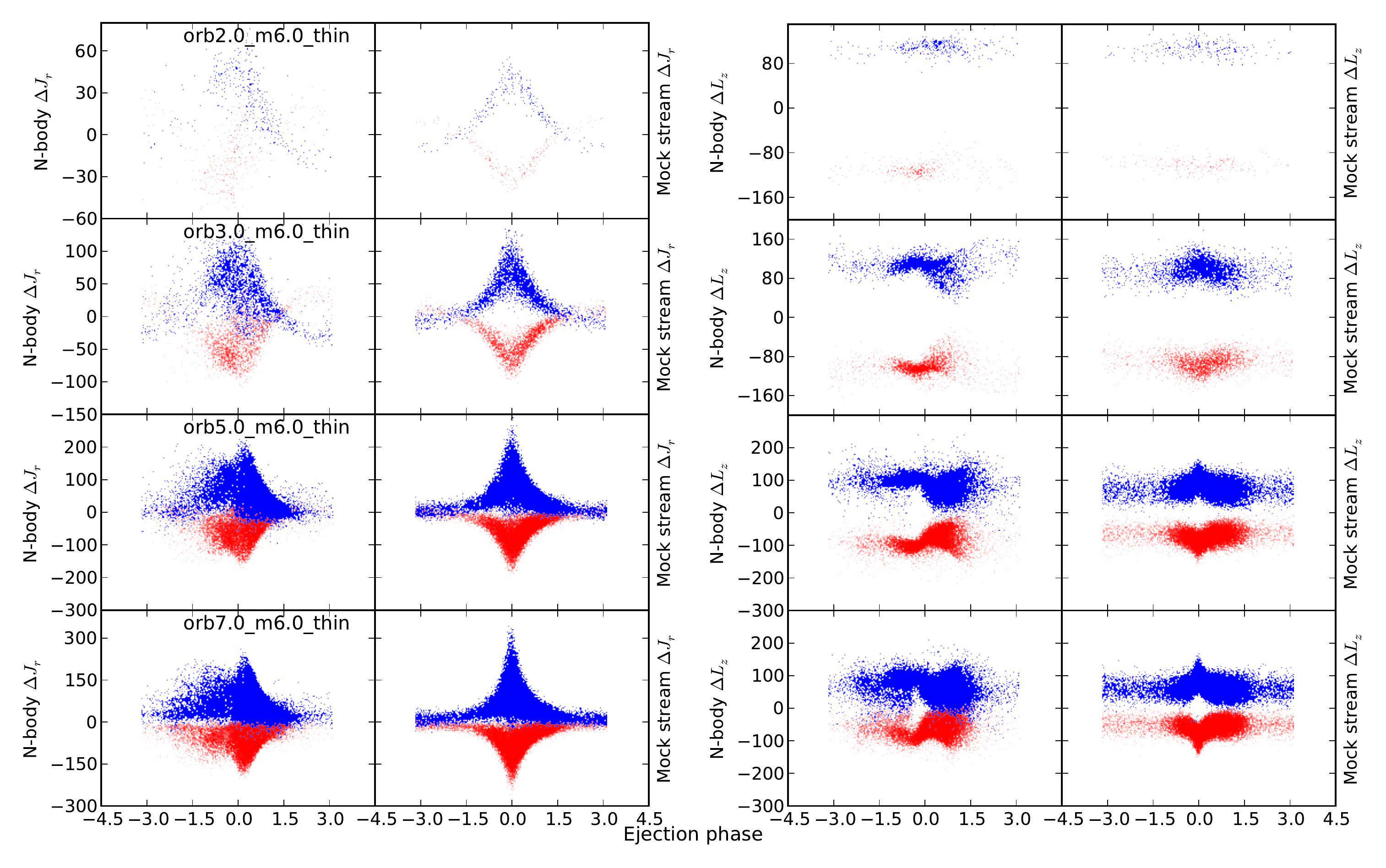}
\continuefloat
\end{figure*}

With these specifications, we can now model the particle emission from
the simulated satellite. We first determine the number of particles to
be ejected between subsequent apocenters, then randomly sample from
equation~\ref{eqn.ejectrate} using the rejection method, which
generates a sequence of ejected particles. The mass in the satellite
is determined by the analytic formula for the cumulative ejected mass.

\subsection{Ejection model}
The last step is to specify the orbital release conditions of 
the ejected particles.  We determine this by comparison with
the final actions and frequencies of the ejected particles
in the simulations as a function of radial phase at ejection.
Since we are using spherical isochrone potential, the azimuthal and vertical frequencies 
are degenerate, so we focus on the behavior in 4-dimensional action-angle space.

We specify the release using equations~\ref{eqn.r}--\ref{eqn.vt}, 
setting $k_\phi=0$, $k_{vr}=0$, and choosing  $k_r$, $k_{v\phi}$ to be constants.
We began by examining the predictions of the \citet{varghese11} and 
\citet{kuepper12} initial conditions discussed previously.  The mean
action and frequency offsets generated by the two initial conditions 
are fairly similar when averaged over a radial cycle, but the
variation of the frequencies with phase 
is much stronger using the K\"{u}pper initial conditions.  
We optimized the choice of constants essentially by eye.
The mean values for the $k$ constants are
\begin{eqnarray}
\bar{k}_r & =&  2.0 \, \mathrm{,} \label{eqn.kr} \\
\bar{k}_{v\phi} & = & 0.3 \; \mathrm{.} \label{eqn.kvphi}
\end{eqnarray}
This is closer to the Varghese conditions than the K\"{u}pper condition, 
the choice of which tends to make the frequencies vary too much with orbital phase.
The optimal parameters appear to vary slightly between simulations, 
but not strongly enough for us to justify a more complex choice.

\begin{figure*}
\twocol{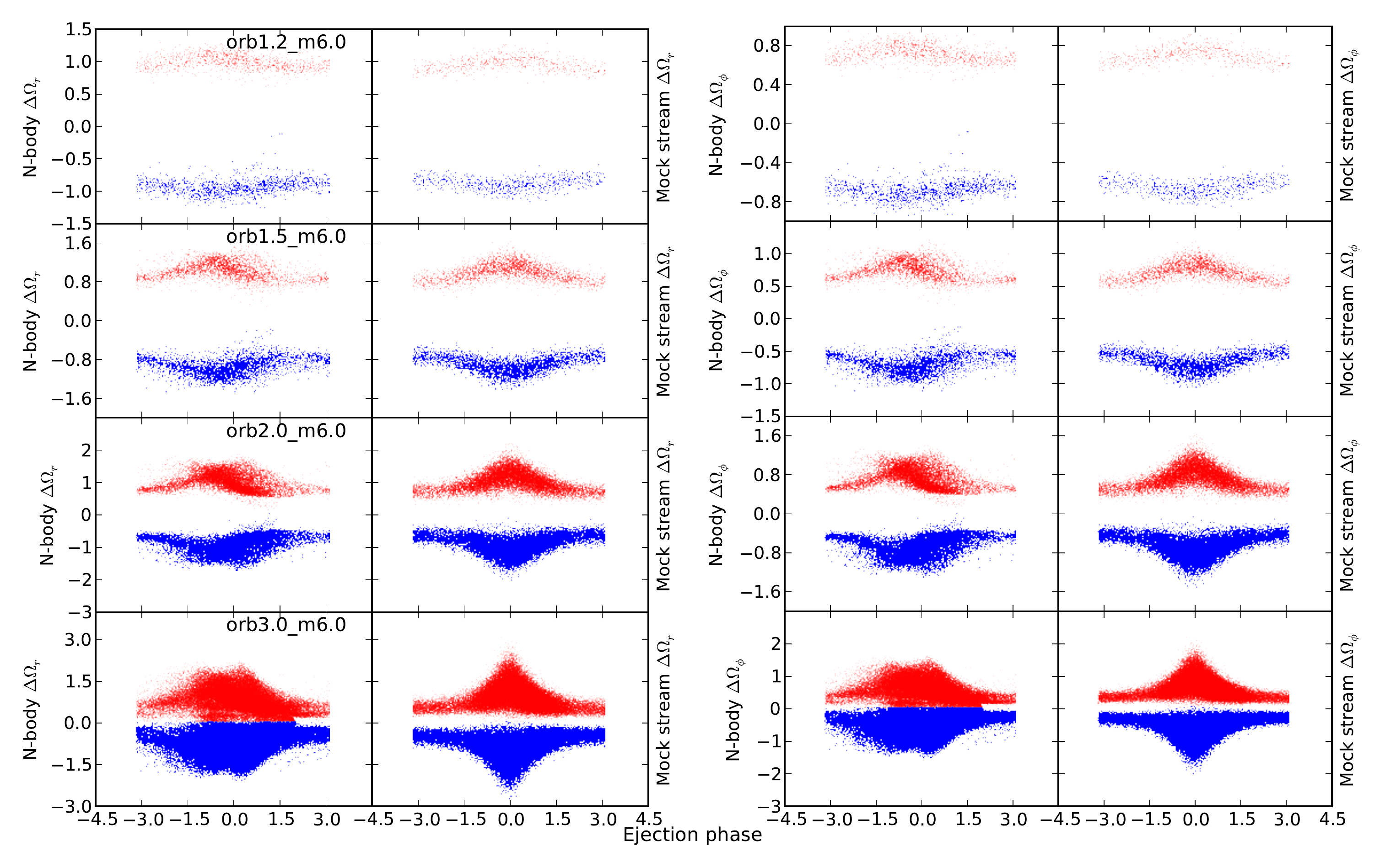}
\caption{
\label{fig.freqs}
Orbital frequencies of stream stars.
First two columns show $\Omega_r$, second two show $\Omega_\phi$.
These are given in units of radians / Gyr.
In each pair of columns, simulation results are on
the left and results from the mass loss model on the right.
}
\end{figure*}

\begin{figure*}
\twocol{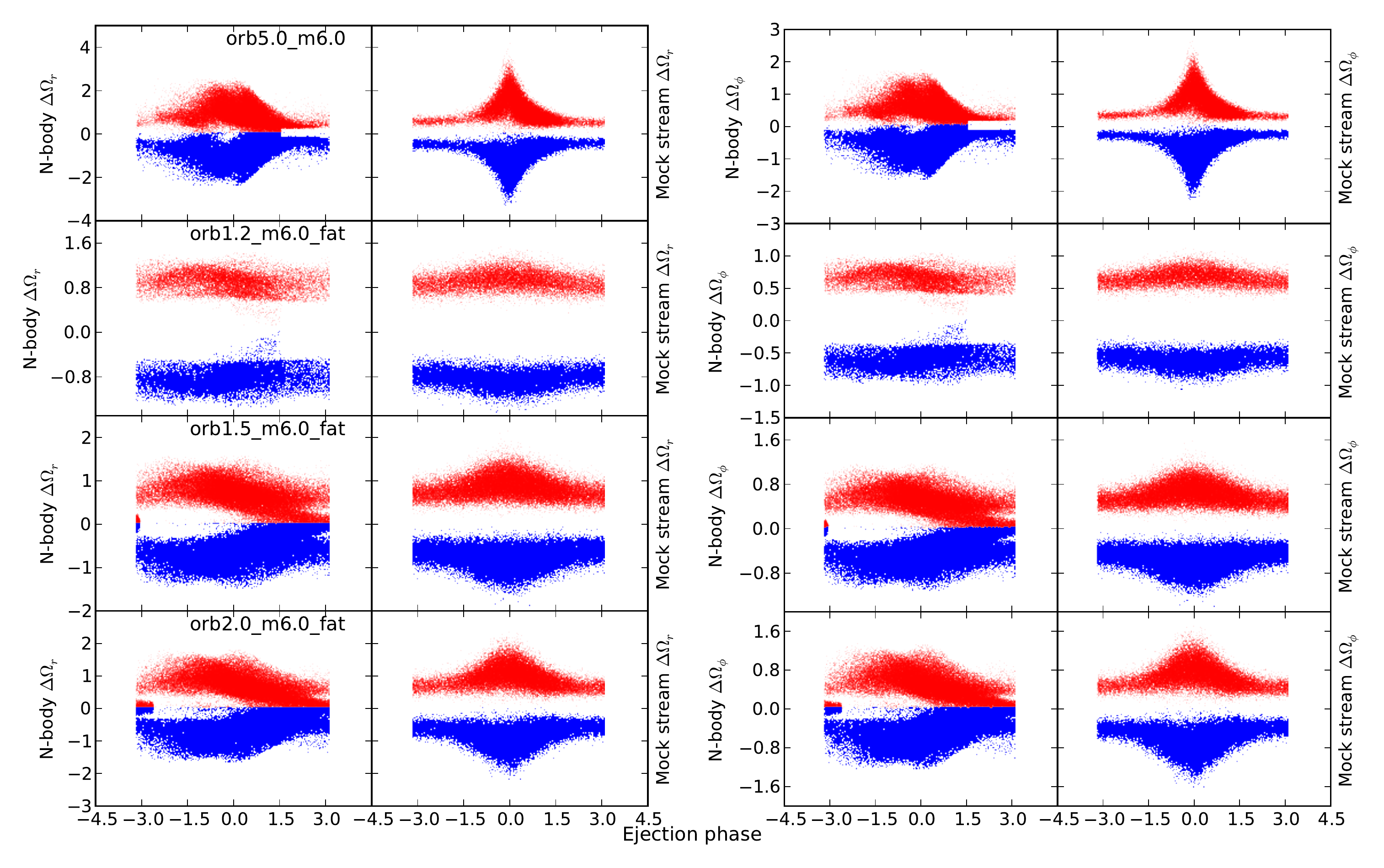}
\continuefloat
\end{figure*}

\begin{figure*}
\twocol{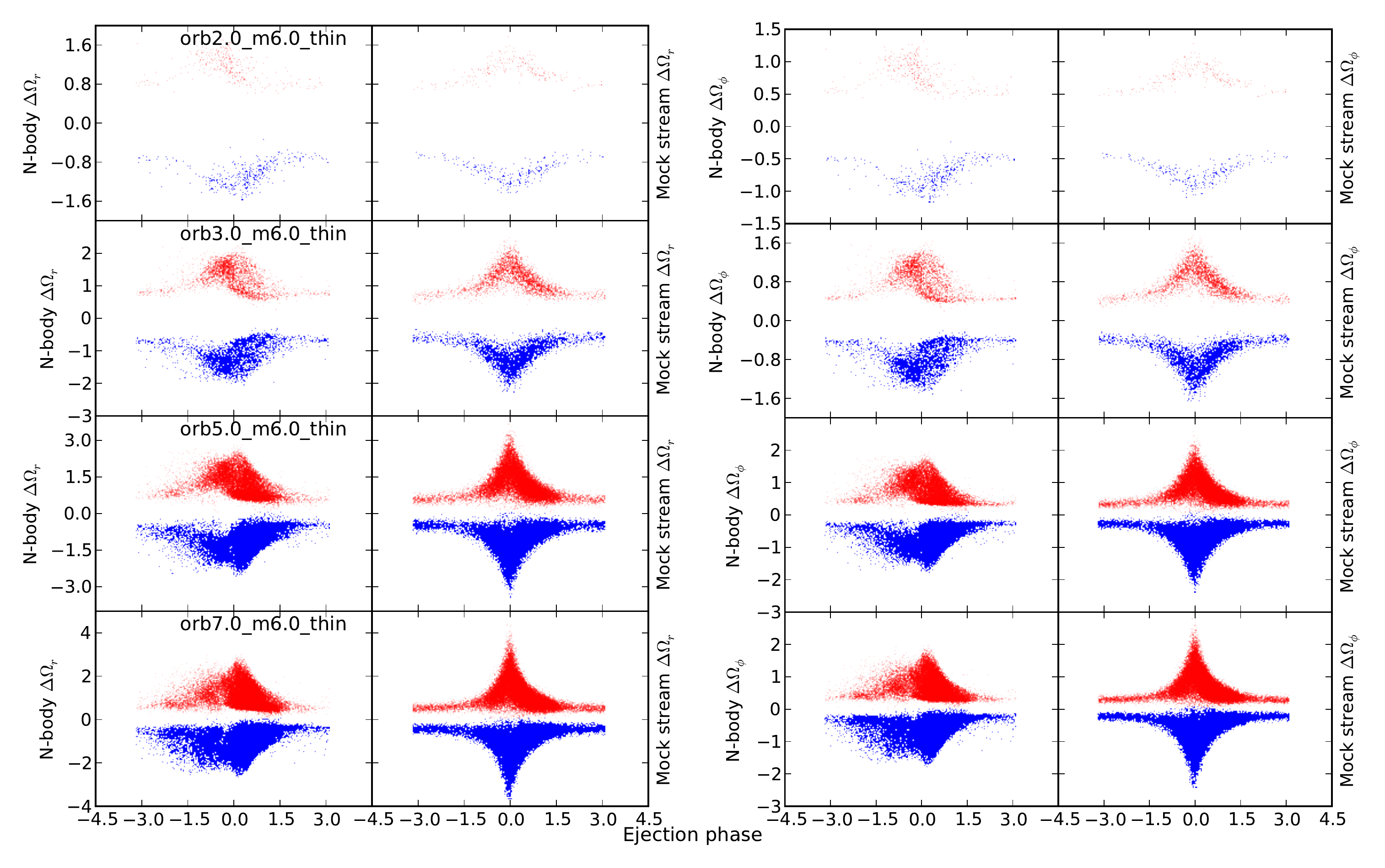}
\continuefloat
% }
\end{figure*}

However, Figures~\ref{fig.actions} and \ref{fig.freqs} show that the {\em dispersion} 
in release conditions is {\em not} constant, but decreases significantly for simulations 
with little mass loss (those on near-circular orbits).
We address this by choosing dispersions $\sigma(k_r)$ and
$\sigma(k_{v\phi})$ of the form
\begin{equation}
\sigma(k_r) = \sigma(k_{vt}) = \min(0.15 f_t^2 \racc^{2/3}, 0.4)
\end{equation}
The initial conditions for the particle-spray models in these and subsequent 
figures include these dispersions.

\begin{figure}
\onecol{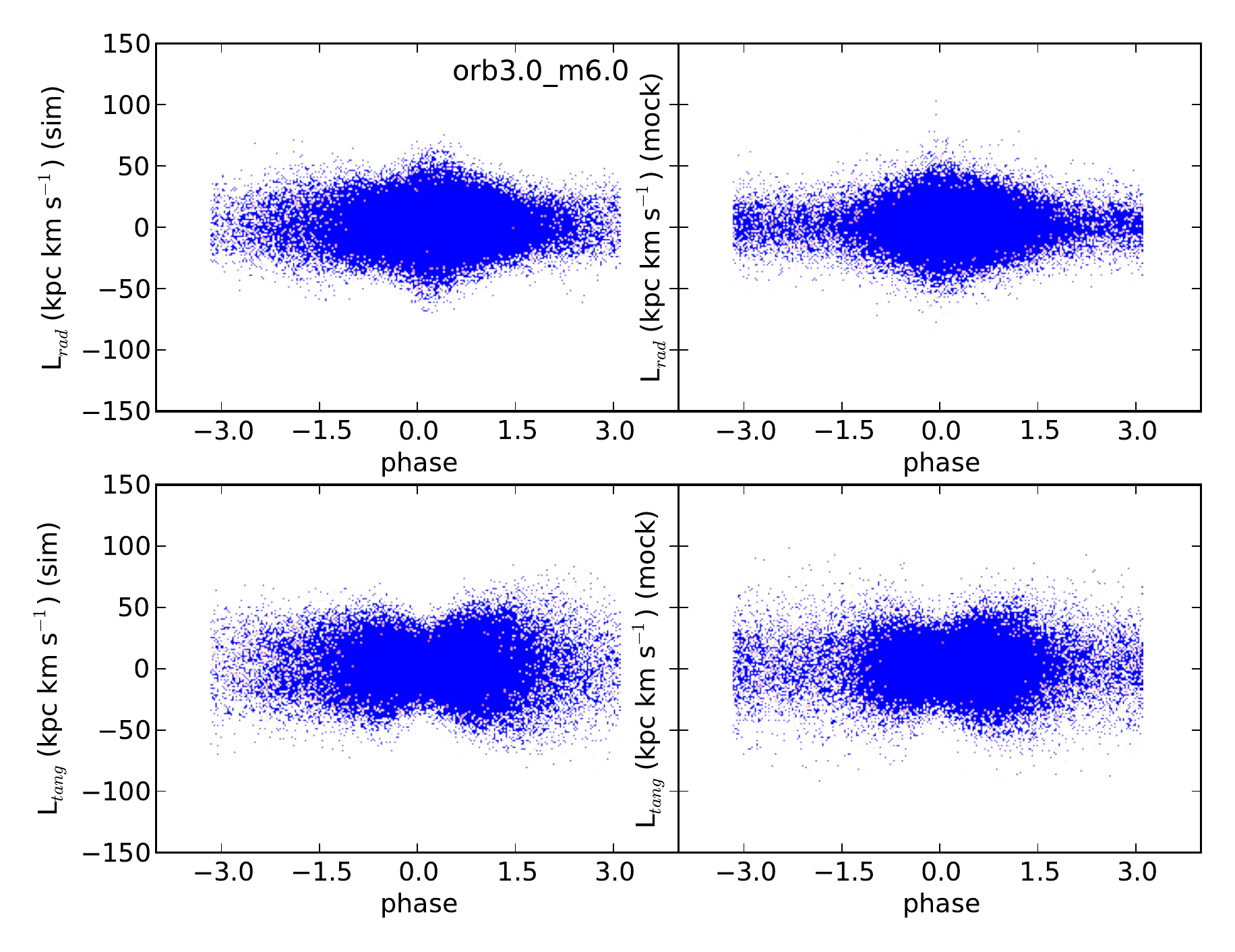}
\caption{
\label{fig.vertical}
Angular momentum components in orbital plane, used as indicators
of off-plane motions of the particles, shown as a function of 
ejected phase.  
First two columns show the component along the radial vector to the
satellite, while the second two show the component along the 
tangential vector.
In each pair of columns, simulation results are on
the left and results from the mass loss model on the right.
}
\end{figure}

We also introduce a random nonzero offset in the off-plane position and velocity.
For spherical simulations, this is necessary if we are to prevent the model
stream from being entirely flat, in contrast to the simulation
results.  
In a non-spherical potential, the vertical extent also thickens 
due to differences between $\Omega_z$ and $\Omega_\phi$.  
We choose
$\sigma(k_z)$ and $\sigma(k_{vz})$ to reproduce the dispersion
in the off-plane components of angular momentum along the 
radial and tangential vectors to the satellite at the time of
ejection.  
In our model given by 
equations \ref{eqn.z}--\ref{eqn.vz}, these two components are
determined by $k_z$ and $k_{vz}$ respectively.  
The choices 
\begin{eqnarray}
\bar{k}_z &=& \bar{k}_{vz} = 0 \; \; \mbox{(by symmetry))} \\
\sigma(k_z) &=& 0.5 \\
\sigma(k_{vz}) &=& 0.5 \label{eqn.sigkvz}
\end{eqnarray}
seems to reproduce the simulation results adequately, at least to the
level we are interested in here (see Figure~\ref{fig.vertical}).   

\begin{figure*}
\twocol{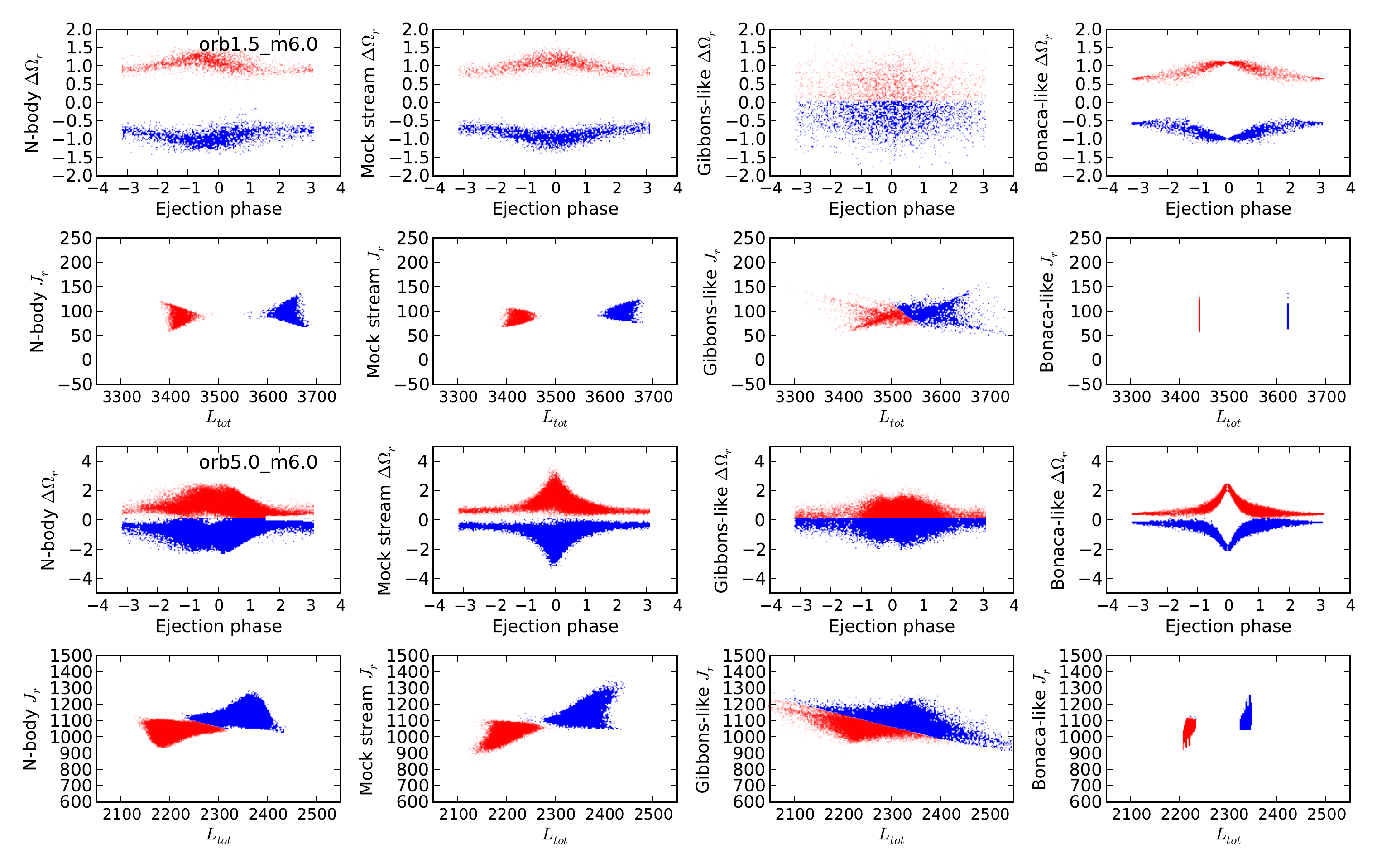}
\caption{
\label{fig.compare}
Comparison of different algorithms.  The first two and second two rows
use the {\tt orb1.5\_m6.0} and {\tt orb5.0\_m6.0} runs respectively.  
The first row for each run shows the radial frequency offset versus ejection phase.
The second row for each run shows the radial action versus the 
angular momentum.  
The first column shows the simulations, the second 
our recipe, the third the 
\citet{gibbons14} recipe (minus the force from the satellite), and the
fourth the \citet{bonaca14} recipe.  
}
\end{figure*}

To generate a mock stream, we randomly sample the values $k_r$,
$k_{v\phi}$ $k_z$, and $k_{vz}$ from Gaussian distributions with the
specified mean and dispersion. We can then determine the positions and
velocities using equations \ref{eqn.r}--\ref{eqn.vz} and
\ref{eqn.kr}--\ref{eqn.kvphi} and then determine the actions
(Figure~\ref{fig.actions}) and frequencies (Figure~\ref{fig.freqs})
for each of our ejected particles. These figures were used to guide
our choice of constants. The action plots of the simulations show a
significant amount of interesting structure in some cases, in contrast
to the particle-spray model which is fairly symmetrical about the
origin and unimodal at any ejection phase. However, the structure in
the frequency plots appears overall less complicated and more similar
to the particle-spray model, though in some cases bimodal structures
or sprays of particles are still evident. The frequencies influence
the stream properties more than the actions, so this is an encouraging
sign for the particle-spray model.

\begin{figure*}
\twocol{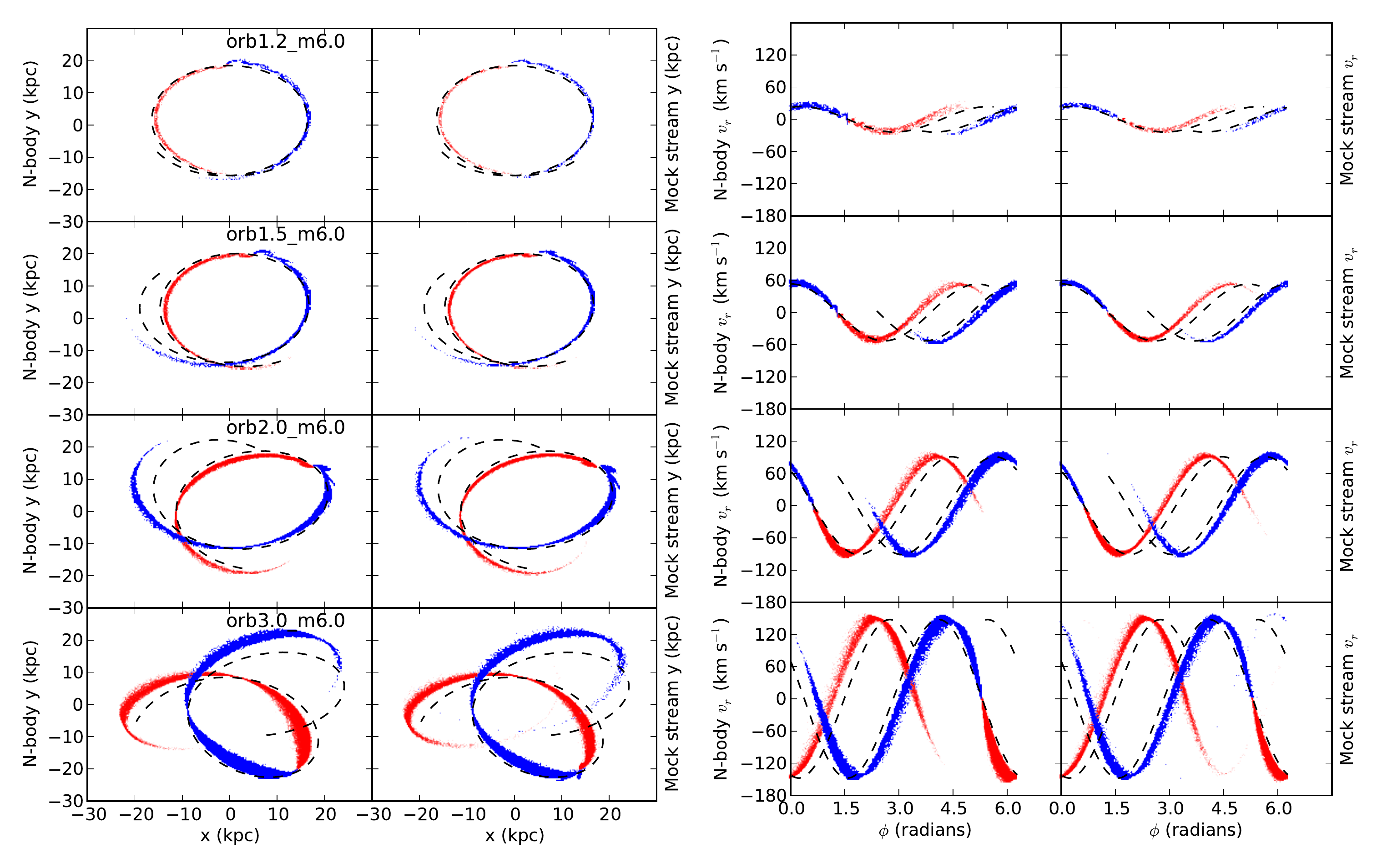}
\caption{
\label{fig.streampos}
Stream position / velocity.
First two columns show a face-on view of the orbit plane,
while the second two show the radial velocity versus the azimuthal angle.
In each pair of columns, simulation results are on
the left and results from the mass loss model on the right.
Streams are shown at the last apocenter timestep at which the
satellite is still intact.
}
\end{figure*}

\begin{figure*}
\twocol{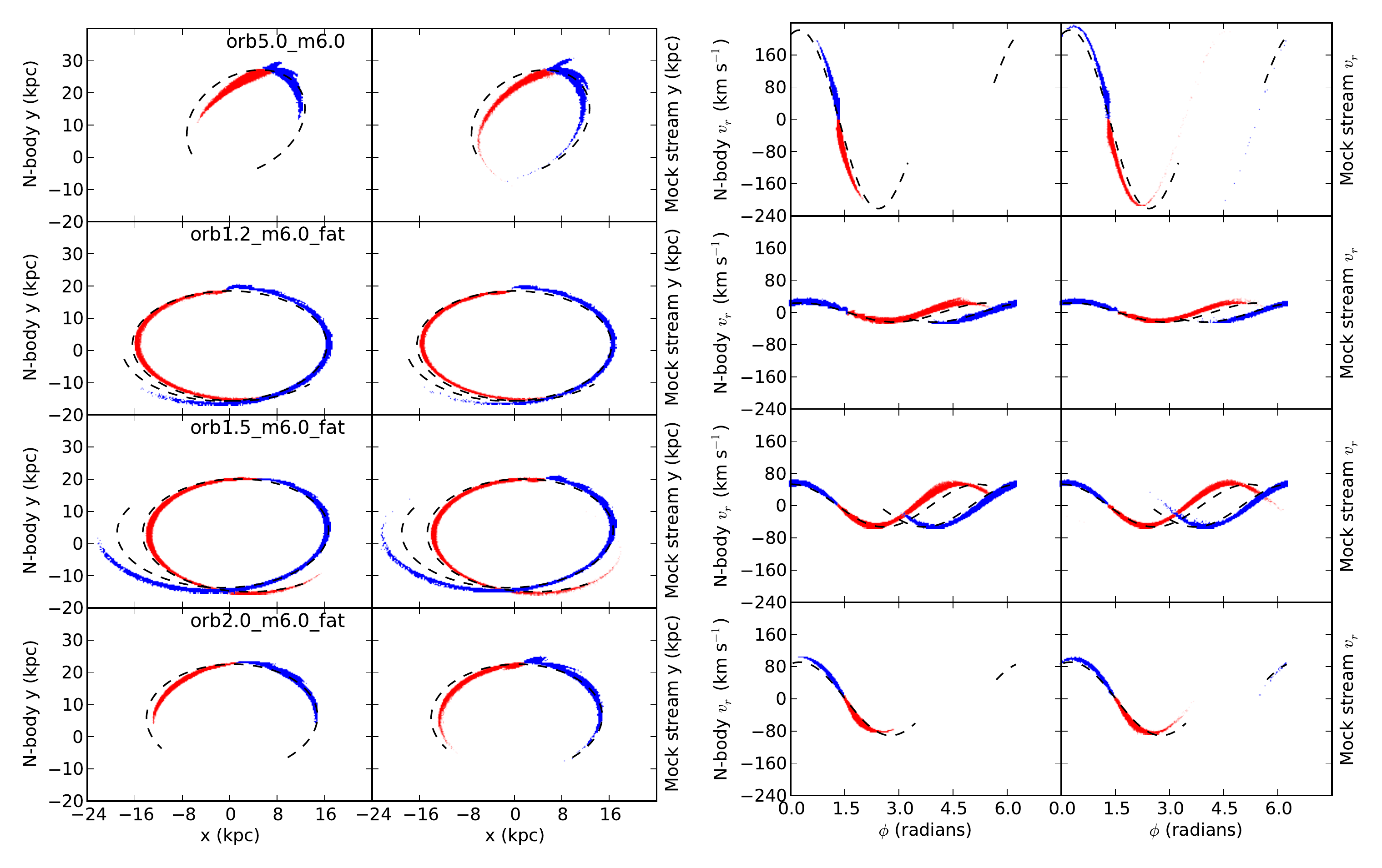}
\continuefloat
\end{figure*}

\begin{figure*}
\twocol{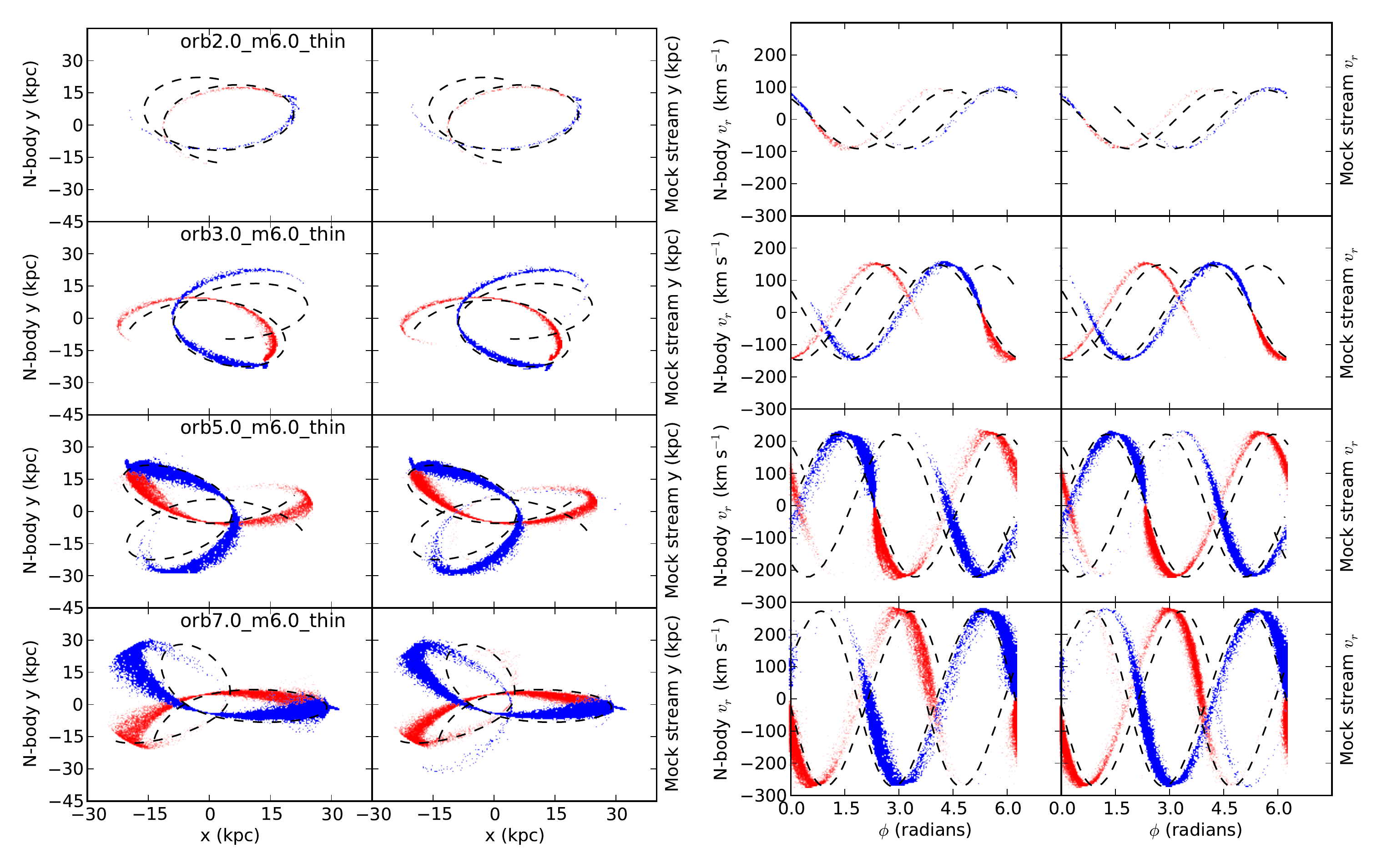}
\continuefloat
\end{figure*}

The correlation between the angular momentum and radial action 
of the stream particles can be seen in the second and fourth rows of 
Figure~\ref{fig.compare}.  Here the first column is the simulation
results, the second is the result of our recipe, while the third and
fourth columns will be discussed later.   It can be seen that our
recipe reproduces the characteristic ``bow-tie'' pattern noted for
this plot in this previous work \citep{eyre11,bovy14}.  

Once we have chosen ejection times and phase-space coordinates
for all the mock stream particles, we can evolve them through time
either by direct orbital integration or by using the action-angle 
formalism.  We use the latter method here because of the ease of
action-angle calculations in the isochrone potential.  We propagate
the angles to the final timestep using the computed frequencies of
each particle.
Finally, we can determine the position and velocity of each
particle by inverting the real-space to action-angle transformation.
The results are shown in Figure~\ref{fig.streampos}.  
The agreement is quite impressive overall, showing deviation
from the orbit by the correct amount and a fairly good agreement
on the length, width, and surface brightness of the stream.  
The main deviations from the simulation can be found very
close to the satellite and in the extreme tails of the distribution,
but a careful inspection is necessary to find these differences.

\section{STREAM GENERATION MODEL IN PRACTICE}
\label{sec.mockstreams}
\subsection{Comparison of stream recipes}
In Figure~\ref{fig.compare}, we compare the results of our recipe to
two other recipes that have appeared recently, which we implemented
to the best of our understanding. The results of the
simulations and our recipe appear in the first and second columns
respectively. The \citet{gibbons14} recipe, which they call the
``Lagrange Cloud stripping'' method, appears in the third column,
though in this case we are {\em not} including an additional force
from the satellite as they found necessary, but simply integrating
particles in the host potential. Also, here we take the velocity
dispersion at particle release to be be $\sigma_v = V_c(r) r_t / r$,
which we found to be obeyed reasonably accurately by measuring the
actual 1-d velocity dispersion of bound particles over our ensemble of
simulations. In the Gibbons approach this dispersion is a free
parameter used in fitting the results of simulations or observations.
The recipe of \citet{bonaca14}, which they call the ``Fast-forward''
method, appears in the fourth column. Both these recipes require an
estimate for the velocity dispersion, but this is taken to be a free
parameter used in fitting in the Gibbons approach.  

For the two simulation
cases shown, the Gibbons recipe results in frequency offsets that are
generally too small. This agrees with their finding that the streams
are too short without an additional force from the satellite. In
contrast, the scale of the offset in the Bonaca recipe is fairly good.
However, the contrast between minimum and maximum frequency
is somewhat too large.

The Gibbons streams have frequency distributions that are too
broad, especially for the first run shown ($r_a / r_p = 1.5$),
so that the leading and trailing streams merge into each
other. The especially narrow simulated frequency distribution in the first
instance is evidence for the need to limit the ejected dispersion for
nearly circular orbits. This plot shows the result when we take
$\sigma$ equal to the velocity dispersion of the satellite. In the
Gibbons recipe, $\sigma$ is actually treated as a free parameter.
So in their approach the optimization should eventually pick a reasonable
dispersion, but at the cost of losing any information obtainable from
linking the release velocity dispersion to the satellite properties.

The Bonaca streams in contrast have a large radial action dispersion,
but almost no dispersion in angular momentum, owing to the inclusion
of dispersion only in the radial velocity. For certain simulations
(see the lowest row in Figure~\ref{fig.compare}), the dispersion is
large enough to throw some stars backwards into the stream emerging
from the opposing Lagrange point. This results in two extra streams
emerging from the satellite in these cases. 
For most simulations, the frequency dispersion is too small at
pericenter, which results in streaky features that are too short
compared to the simulations.
In summary, our recipe seems to improve the overall agreement with
simulations compared to previous recipes for the cases examined here,
though the differences will be more visible in some cases (mainly
those with particularly small or large eccentricities) than in others.
The general similarity, along with the agreement with simulations
we demonstrated in section~\ref{sec.model}, 
supports the general approach of particle-spray
methods as used here and in the work of \citet{gibbons14} and
\citet{bonaca14}.

\begin{figure*}
\twocol{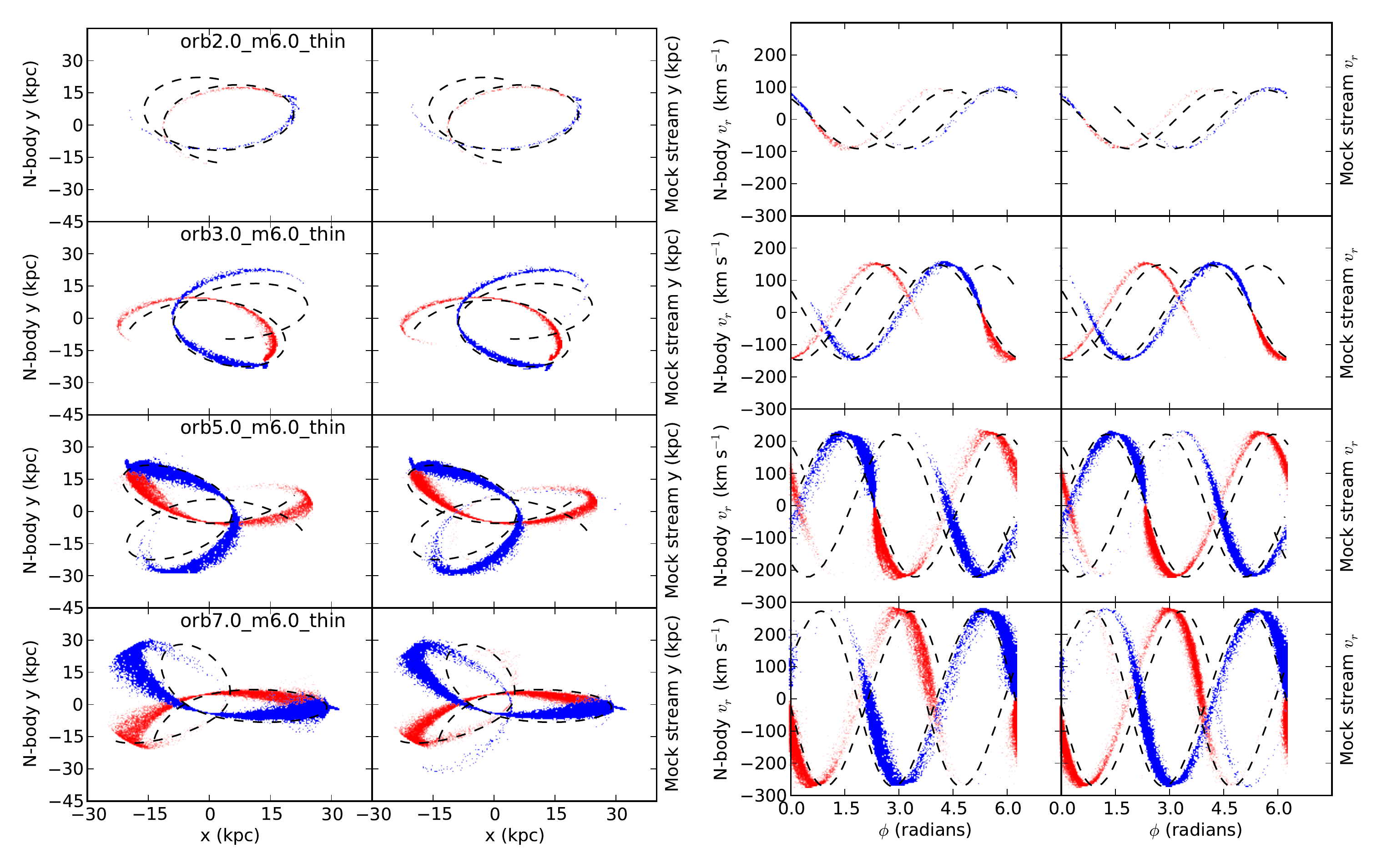}
\continuefloat
\end{figure*}

\begin{figure*}
\twocol{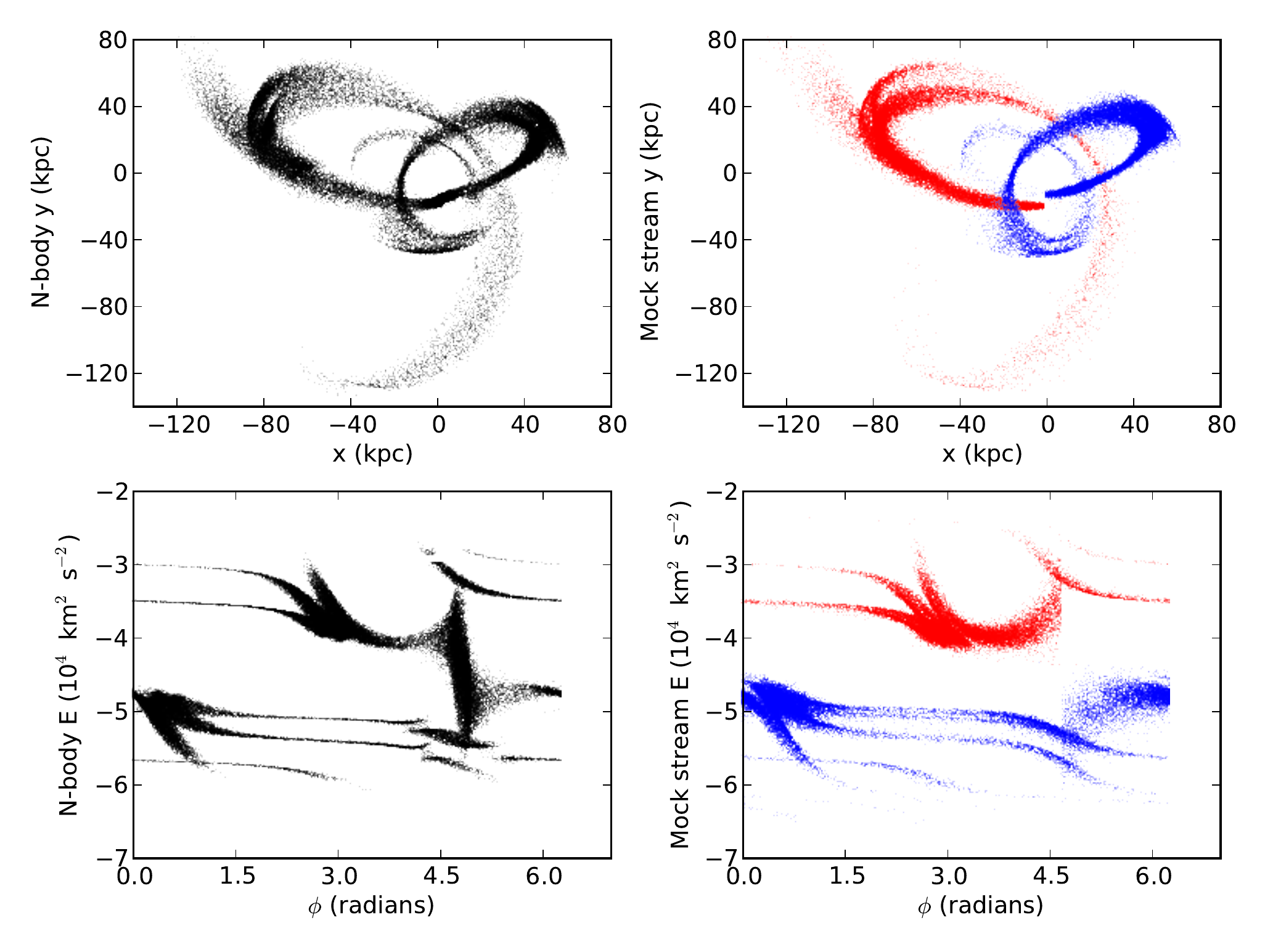}
\caption{
\label{fig.sagrun}
\nbody\ simulation roughly emulating the Sagittarius stream (left column), 
following \citet{gibbons14},
and particle-spray model thereof (right column).  
The top row shows particle positions. The bottom row shows the 
particle energies versus azimuth; streams from different pericenter
passages are particularly well separated in this representation.  
The progenitor appears at $x = 0 \kpc$,  top left vertical streak in 
The complex structure of
the simulated stream is well reproduced by the particle-spray model.
}
\end{figure*}

Turning now to the overall ensemble of simulations, the model streams
shown in Figure~\ref{fig.streampos} agree extremely well with the
simulation results. The width, length, and substructure within the
streams are all quite well reproduced. The most common deficiency is
an excessively long tail of debris, which stems in part from our use
of simple Gaussian distributions of the space and velocity offsets
rather than a more sharply truncated distribution.

\subsection{Different host potentials}
To extend the test of the method beyond the parameter range where it
has been calibrated, we run an \nbody\ simulation based on the one
shown in Figure~2 of \citet{gibbons14}.  This simulation uses
a different potential and a much more massive satellite than those
we have examined so far.
The host potential for this run is a spherical
Navarro-Frenk-White profile, with a mass of $7.5 \times 10^{11} \msun$
contained within a radius of $185.41 \kpc$, and a 
scale radius of $9.27 \kpc$ yielding a concentration of 20.  
This equates to a virial overdensity of 180 with
respect to the critical density for a Hubble constant of $H_0 =
75 \kms$.    We use a $W=3$ King model of $6.4
\times 10^8 \msun$ and outer radius $4.75 \kpc$ 
We set this moving in the $x$-$y$ plane with
coordinates $x_0= -2.327 \kpc$, 
$y_0 = 70.772 \kpc$, $v_{x0} = -78.71 \kms$, and $v_{y0} = -2.59 \kms$.
The resulting orbit has apocenter $70.8 \kpc$ and pericenter $17.8 \kpc$,
and combining this orbit with the satellite properties leads to
a tidal factor $f_t = 0.8$.  

The simulation results are shown in the left column of
Figure~\ref{fig.sagrun}, at a time $4.14 \Gyr$ into the run. At this
point the satellite is just past the fourth pericenter, so three
separate streams from previous pericentric passages are clearly
visible in both the leading and trailing streams. The spatial 
distribution of the particles
is similar to that of \citet{gibbons14}. 
The different streams separate particularly well in the lower row, which
shows the particle energies plotted versus azimuth.   In the spatial
distribution, these streams separate most clearly at apocenter of the 
radial loops.  This substructure clearly shows the action-angle tilt 
which is expected to dominate for a young stream (only 3.5 radial
oscillations), as discussed above.

Results from our particle-spray model are shown in the right-hand
column. We have performed the required orbit integrations using the
{\tt galpy} package written by Jo Bovy. Clearly, the complex structure
of the stream is well reproduced by our model. When overlaid, the
multiple components of the leading and trailing streams match the
\nbody\ model nearly exactly. Some slight differences are visible,
notably gaps at two different azimuth values in the leading stream of
the simulation and one in the leading stream in the energy plot. One
of these gaps is also visible in the spatial distribution, and results
in a ``C''-shaped feature separated from the rest of the leading
stream. These features result from the satellite crossing its own
stream, as made inevitable by the perfectly spherical potential used
here, and creating a gap in the manner studied by \citet{carlberg13}.
Since our model does not include the satellite potential, these gaps
do not appear in the right-hand column. The model stream is somewhat
longer and in some places narrower than the \nbody\ results. Also,
differences are easily apparent near the progenitor satellite, both
from the progenitor itself (not represented in the particle-spray
model) and from particles that have been stripped but have yet
to acquire their full action offset as they curve away from the
progenitor.  Overall, however, the agreement is extremely good.

\begin{figure*}
\twocol{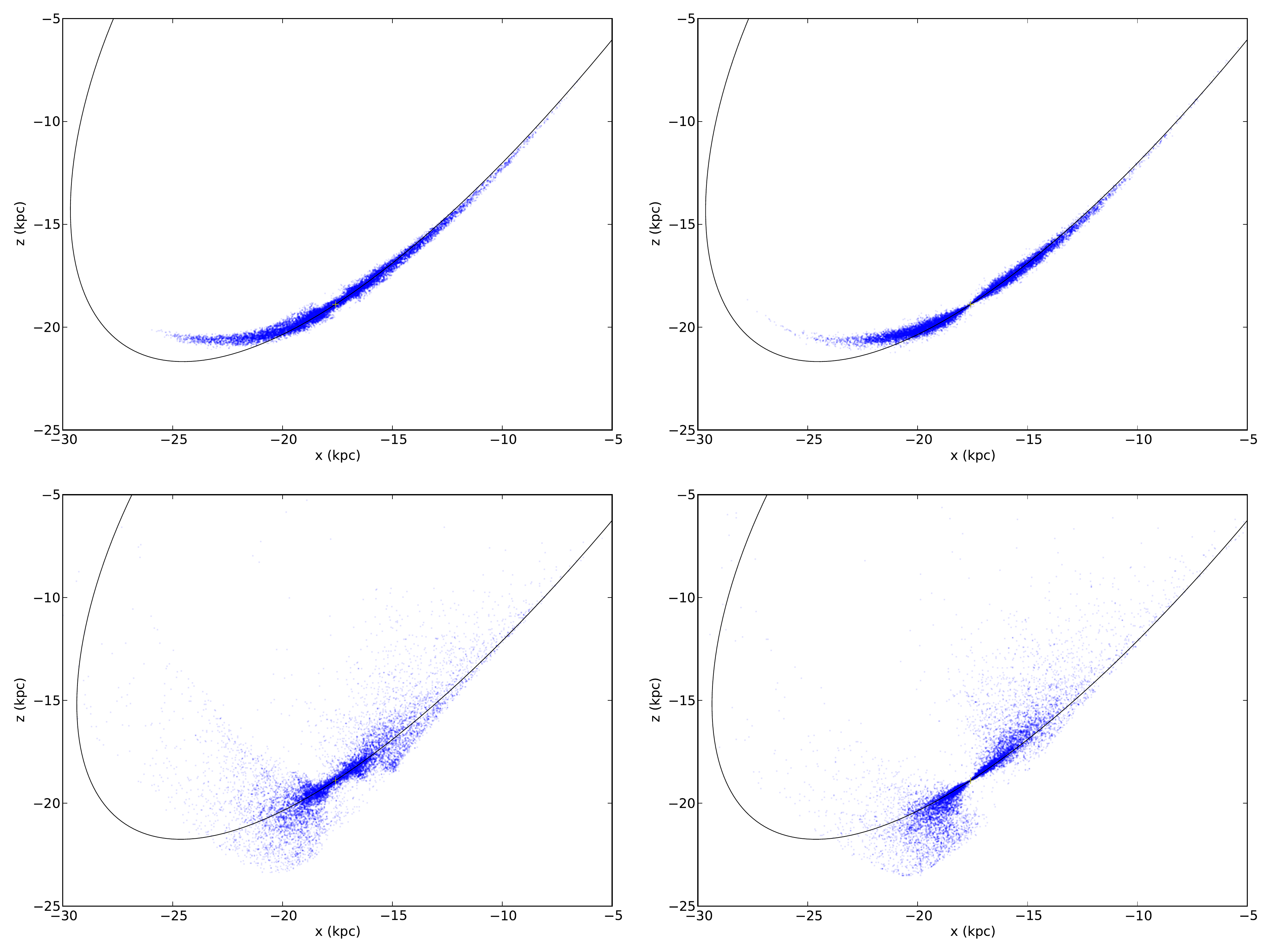}
\caption{
\label{fig.chaos}
Simulations using a bulge-disk-halo potential, as discussed in the text
Left columns: \nbody\ simulations.  Right columns: particle-spray models.
Upper row: regular orbit with $q=0.57$.  Lower row: 
apparently chaotic orbit with $q=0.60$.  
}
\end{figure*}

Figure~\ref{fig.chaos} shows the performance of the particle
spray recipe in a compound bulge-disk-halo potential. Two orbits are
shown, one on a regular orbit and one on a mildly chaotic orbit. The
satellite in this test is located at a position 
$x_0=-17.59 \kpc$, $y_0=-10.55 \kpc$, $z_0=-18.89 \kpc$ and
velocity $v_{x0}=-119.8 \kms$, $v_{y0}=24.36 \kms$, 
and $v_{z0}=-83.12 \kms$. We initialize the satellite with
mass $4 \times 10^4 \msun$ and tidal factor $f_t=0.7$. 
The orbit is integrated backwards from the given
point for $t_d= 2694\Myr$, and then the particle-spray and \nbody\
simulations are evolved to the present day. In the fixed potential
used here, the bulge is a Hernquist profile with mass 
$3.4 \times 10^{10} \msun$ and radius $0.7 \kpc$.
The disk is a Miyamoto-Nagai disk with mass $10^{11} \msun$,
$a=6.5 \kpc$, and $b=0.26 \kpc$. The halo
is an oblate logarithmic halo with 
$\phi(R,z)=V_h^2 \ln (R^2 + (z/q)^2 + d^2)$, 
$V_h=115 \kms$, and $d=12 \kpc$. 
The only difference between the orbits used in the top and
bottom rows is that the top uses $q=0.57$, and the bottom $q=0.60$.

The thickness of the streams in the top and bottom rows differs markedly,
for both the particle-spray models and N-body simulations.
This appears to be due to the nature of the progenitor orbits. The
orbit in the top case is regular, as we verified using a Poincar\'{e}
surface of section plot. In contrast, the surface of section for the
bottom case is slightly thickened, indicating a small degree of chaos.
If we further increase the
flattening parameter to $q=0.63$, the orbit becomes regular again, and
in tandem the stream becomes narrow again so that it much more closely
resembles the top panel. (Admittedly, the flattening in all of these halo potentials 
is perhaps unrealistically high, but chaotic orbits can be found in more
realistic potentials as well, as in \citealp{hunter05}.)  It is worth
noting that in this case, the thickening of the stream arises within less than
five radial oscillations.  Long integration times are
apparently not required to see the influence of chaos on tidal streams.

The results in Figure~\ref{fig.chaos} demonstrate we can obtain reasonable 
results from the particle-spray model, even when the progenitor orbit is
irregular, though some areas of disagreement can be found in both
rows.  In fact, having first noticed a highly thickened stream 
in a similar orbit and potential using the particle-spray model,
we suspected a bug in our code until the \nbody\ model
demonstrated the same behavior. For the spray model to work on 
a chaotic orbit, the particle coordinates must be evolved by direct orbit
integration rather than by using an action-angle formalism.

The fraction of orbits exhibiting chaotic behavior is generally small,
though nonzero, in the idealized axisymmetric models often used in
stream modeling \citep{hunter05}. However, the fraction of orbits
demonstrating chaotic behavior in {\em realistic} galaxy potentials,
which include triaxiality, radial shape dependence, time dependence,
and substructure, may well be higher. Also, real galaxy potentials and
hence the orbits within them evolve with time. In principle, orbits
may sweep through chaotic regions, puffing up their streams. This
would raise the fraction of tidal streams affected by chaotic behavior
beyond the fraction of orbits that are currently chaotic. 

In this last set of simulations, we chose the parameters such that the
satellite remains partly intact until the end of the simulation. If
the satellite is more diffuse, it comes apart earlier in the \nbody\
run than in the particle-spray model.  Presumably 
this is due to disk shocking,
though we have not investigated the cause in detail. The particles are
also spread more evenly along the stream than in the particle-spray
model, where there tends to be a central hole. These discrepancies
join those mentioned earlier as the main discrepancies with the
\nbody\ results: differences in the tails of the particle
distributions, inaccuracy in the satellite mass as a function of time,
and inaccuracy of the streams near the satellite Lagrange points. However,
overall the agreement between these two methods is impressive,
suggesting the stream recipe is suitable for fast modeling of streams
in a Bayesian context.

\section{DISCUSSION}
\label{sec.discussion}
We have presented a recipe for modeling
tidal streams without the cost of a full \nbody\ simulations. 
The typical computation times are a few minutes, as opposed
to a typical computation time of six hours for the \nbody\ runs
in Table~\ref{table.sims}.
This recipe has many similarities with recent suggestions by
\citet{gibbons14} and \citet{bonaca14}, but differs in several
respects. Here we discuss some aspects of our recipe in the context of
these and other methods.

Our method clearly reproduces the differences between the stream and
the orbit that were reviewed in Section~\ref{sec.actionangle}. As
discussed in that section, the stream behavior is determined primarily
by the distribution of orbital frequencies (or equivalently the
actions) upon release. Even though we have made simplifying
assumptions about the way the stars are released, and neglected the
subsequent influence of the satellite, the behavior of streams is
still remarkably faithful to the \nbody\ results.

Our model also reproduces the dispersion along various dimensions of
the stream fairly well. Streakline techniques such as those of
\citet{varghese11} or \citet{kuepper12} by design do not model this
dispersion. Therefore, they fail in particular to reproduce the
portions of the stream furthest from the satellite, which are dictated
by the extremes of the frequency distribution at the first disruptive
pericenter. They similarly fail to reproduce the extent of any other
lobes or streaky features within the stream. Thus while they are
certainly easier to compute, the folded streaklines evidenced by
\citet{kuepper12} are not necessarily easier to fit to the stream than
a full particle distribution. In addition, the length and width of the
stream contains useful information on the satellite mass and overall
duration of the disruption process. These parameters are in turn
valuable in obtaining a more accurate fit to the orbit and the
potential.

Substructure in streams provides important constraints,
as described in section~\ref{sec.substructure} and further illustrated
in section~\ref{sec.mockstreams}.
A crucial point in modeling these streams is that the actions
and frequencies of released stars must vary over the radial cycle. In
the recipe of \citet{bovy14} for example, the distribution of actions
and frequencies is assumed constant. This type of model is useful in
modeling old streams where the substructure overlaps, but is unable to
reproduce the complex spatial pattern seen in for example
Figure~\ref{fig.sagrun}.  The ejection recipe in our model naturally
reproduces the variation in orbital properties.  In principle, this
variation can be incorporated into action-angle stream models
as well, which would be useful in obtaining large particle numbers 
(or accurate stream distribution functions) at a fixed computational
cost.  

While an action-angle approach has important advantages, irregular
orbits will pose a problem for this approach. For these orbits,
directly integration of the orbits as we have done here seems the only
viable method. The results shown in Figure~\ref{fig.chaos} suggest
that chaotic orbits result in puffed-up streams within a few orbital
times. Further investigation into the effects of chaos on realistic
galactic potentials and tidal streams within them seems warranted.

Our model has a varying emission rate peaking near pericenter, 
in contrast to the constant emission used by \citet{bonaca14}
and \citet{bovy14} among others.
We have tested the effect of a uniform emission rate versus the
simulation in Figure~\ref{fig.sagrun}.  The effects are more
subtle than those of the oscillating actions and frequencies;
the substructure is still easily visible.  However, the influence
of the emission rate is still evident in the abundance of particles
between the clearly separated streams, which fill in the distinct shells
seen at apocenter of the radial loops and blur the separate
streams visible in the energy plot.  Thus this effect too is important
in modeling streams with visible substructure. 
 
Our model also explicitly takes into account the decrease of the satellite
mass with time. This improves upon the treatment in the models of
\citet{gibbons14} and \citet{bonaca14}, as well as the earlier
streakline models, where the mass is assumed fixed. Models with
multiple interactions can be modeled fairly accurately with our simple
assumptions, as long as the satellites and orbits resemble those used
in our tests. For less disruptive encounters, more careful modeling of
the tidal excitation is probably necessary. Very disruptive encounters
will require a criterion for total disruption and a prescription 
for the particles ejected at that point.  Including a disk
in the host potential could lead to disk shocking and a more
rapid mass loss than specified by our current recipe.
Finally, different satellite profiles will probably require a
recalibration of the mass loss prescription. As discussed above,
previous work has shown that satellites with various density profiles
evolve in a predictable manner during moderate rates of tidal
stripping, so this recalibration should be possible as long as the
satellites are dynamically hot systems. The mass evolution in turn
affects the actions and frequencies of released particles, and
ignoring it in the case of Sagittarius for example may
lead to significant biases.

When compared to the \citet{bonaca14} method, two more differences
stand out. First, they release stars at the angular velocity of the
satellite, which results in a larger oscillation of actions and
frequencies over the radial cycle than in our equations~\ref{eqn.vt}
and \ref{eqn.kvphi}. Second, they blur the release velocity of 
the satellite by the velocity dispersion in the satellite,
which in many cases is a
larger dispersion than we use. In some cases their assumptions work
well, but in cases with small or large eccentricities we find that
they can produce streams that are too smeared out in azimuth or
radius. The streams in \citet{bonaca14} are both produced and analyzed
with the same method, so it is unclear whether these assumptions have
much effect on the larger conclusions in their paper.  
Still, these differences
are only strongly evident in certain situations.  Overall, our results
strongly validate their general approach of modeling particle sprays
using orbits in the host potential alone.

The method of \citet{gibbons14} is somewhat different, in that they
include an ongoing force from the satellite in calculating the
particle evolution. It thus closely resembles a standard restricted
\nbody\ method. (A simple alternative to their method is to run a
restricted \nbody\ simulation after first removing the central particles of the
progenitor, because they remain bound to the satellite and add
nothing but computation time.) In their runs, between a quarter and a
half of the released particles are recaptured by the satellite and
later released in bursts at pericentric passage. Our method has the
advantage that we do not need to calculate the evolution of these
recaptured particles, which require short timesteps compared to the
freely moving stream particles. \citet{gibbons14} find that the force
from the satellite must be included.  However, we find that it can be
omitted with an appropriate choice of release constants
(equations~\ref{eqn.kr}--\ref{eqn.sigkvz}); the time-integrated 
effect of the progenitor is encapsulated by our larger action offset. 
One can then
use a simpler, general-use package for calculating the orbital
evolution, rather than more specialized codes. In our case we have
used the {\tt galpy} package which contains a large assortment of
potentials when orbit integrations are required. 

Another difference is that our method prescribes the rate of particle
release as a function of orbital phase. While this form may not be
completely accurate, it is unclear whether the alternative of using
recaptured particles to generate bursts biases the results in some
way. Finally, the velocity dispersion of released particles is
prescribed by our recipe, and tied to the satellite mass and other
parameters of the problem, where in \citet{gibbons14} it is taken to
be a free fitting parameter. Our method thus discards less
information, in principle increasing the statistical power of the fit.
The main advantage of the \citet{gibbons14} method is the automatic
inclusion of multiple interactions between the satellite and its
stream, which can create stream gaps like those visible in
Figure~\ref{fig.sagrun}.  Also, at very large satellite masses the
effect of the satellite will not converge within an orbital wrap and
this will alter the morphology of the tidal stream 
\citep{choi07}, making explicit inclusion of the satellite force necessary.

Of course, one may question to what extent the idealized \nbody\
simulations used in this paper reflect the cases expected in reality.
\citet{bonaca14} describes an interesting experiment, where streams
were generated in both fixed potentials and in a cosmological
simulation of a halo, and then fitted to constrain a parameterized
model of the dark matter halo. While the fixed potential streams
consistently returned accurate results, the cosmological halo streams
yielded highly biased or imprecise results in many cases.
Bonaca et al.\ suggest the differences arise from gradual evolution
in the potential, interaction with subhalos, and deviations of the
true potential shape from the fitted form.  We also note the effect of
chaotic orbits, as in Figure~\ref{fig.chaos} above, may play a role
in these differences.  There are some reasons to think the results
of Bonaca et al.\ are too pessimistic about the scientific yield
from studying tidal streams.
Neither of the two ``streams'' displayed from their much larger sample 
particularly resembles prominent Milky Way streams in 
the cosmological case, while one of them does not look
much like a stream in the fixed-potential case either. 
We suggest that in the outer halo, progenitors that are more massive
than the globular-like systems they employ (mass of $2 \times 10^4
\msun$) are necessary in order to generate streams that yield
robust measures of the potential.
(The fixed-potential case does yield good results, but this presumably
depends on the assumption of perfect six-dimensional observational
data.)   This would align with earlier work using more massive progenitors
\citep{siegalgaskins08}, which found only minor effects of substructures 
on tidal streams. 
Still, the analysis of \citet{bonaca14} raises interesting
challenges to the project of stream fitting, and further work on how
to interpret and combine results from individual stream fits is
clearly required.

To the challenges facing stream models, we may add the issue of
dynamical friction. Stream models are often initialized with a
satellite at a particular place and time, usually a particular
pericentric passage. Dynamical friction is frequently invoked as a way
of bringing this satellite onto its destructive orbit, but it is then
often neglected in the actual calculation of the stream, or else
applied only to the satellite but not the stream.  Full \nbody\
simulations with a live host can implement dynamical friction
accurately. However, this technique is extremely expensive, since high
particle numbers are necessary to suppress the effects of particle
noise.  Thus at this point, a good technique for incorporating
dynamical friction consistently in stream models has yet to be
introduced.

\section{CONCLUSIONS}
\label{sec.conclusions}
In summary, we have presented and tested a recipe for modeling tidal
streams as a collection of particles, released and evolved in the host
potential {\em without} the added influence of the progenitor
satellite. This recipe includes a prescription for calculating the
mass lost from the progenitor, and thus the mass in the tidal stream
as well. The mass loss and orbital properties of the debris both
oscillate with the satellite position, producing substructure within
the tidal stream.  The recipe assumes hot, compact one-component
progenitors, and will require adaptation for more complicated cases,
or orbits very different from the moderately disrupting cases examined here.

The intended use of this recipe is for Bayesian sampling, which
requires a very large number of trial models to obtain accurate
constraints on parameters.  The particle-spray method 
adds one more arrow to the
quiver for those seeking to use tidal streams to obtain physical
insights. The observational situation has been developing rapidly.
Numerous streams have been found in the Milky Way using the Sloan
survey, and in M31 with the PAndAS survey. New techniques have
revealed tidal streams in more distant galaxies with even modest-sized
telescopes, and kinematic tracers have been demonstrated in some of
these streams as well. Forthcoming results from Pan-STARRS and Gaia,
among other surveys, should transform the field of tidal streams into
a high-precision industry.  We expect particle-spray modeling such
as that tested here to be one of its chief tools.  

\section*{ACKNOWLEDGMENTS}
We thank Tom Quinn and Joachim Stadel for the use of {\tt PKDGRAV}, 
and Josh Barnes for the use of {\tt ZENO}. We thank
Jo Bovy for making public the {\tt galpy}
package and assisting with its use,
We also thank Jo Bovy and Aaron Romanowsky
for interesting discussions of tidal stream physics and observations. 
MAF, MDW, and SH
acknowledge support by NSF grant AST-1009652
to the University of Massachusetts.  MDW also acknowledges
the support of NSF AST-0907951.  

\footnotesize{
\bibliographystyle{mn2e}
\bibliography{m31,eject}

\begin{thebibliography}{}

\bibitem[\protect\citeauthoryear{{Aarseth} \& {Heggie}}{{Aarseth} \&
  {Heggie}}{1998}]{aarseth98}
{Aarseth} S.~J.,  {Heggie} D.~C.,  1998, \mnras, 297, 794

\bibitem[\protect\citeauthoryear{{Belokurov}, {Zucker}, {Evans}, {Gilmore},
  {Other} \& {Authors}}{{Belokurov} et~al.}{2006}]{belokurov06}
{Belokurov} V.,  {Zucker} D.~B.,  {Evans} N.~W.,  {Gilmore} G.,  {Other} S.,
  {Authors} D.~M.,  2006, \apjl, 642, L137

\bibitem[\protect\citeauthoryear{{Binney} \& {Tremaine}}{{Binney} \&
  {Tremaine}}{2008}]{bt2}
{Binney} J.,  {Tremaine} S.,  2008, {Galactic Dynamics: Second Edition}.
Princeton University Press

\bibitem[\protect\citeauthoryear{{Bonaca}, {Geha}, {Kuepper}, {Diemand},
  {Johnston} \& {Hogg}}{{Bonaca} et~al.}{2014}]{bonaca14}
{Bonaca} A.,  {Geha} M.,  {Kuepper} A.~H.~W.,  {Diemand} J.,  {Johnston} K.~V.,
     {Hogg} D.~W.,  2014, ArXiv e-prints

\bibitem[\protect\citeauthoryear{{Bovy}}{{Bovy}}{2014}]{bovy14}
{Bovy} J.,  2014, ArXiv e-prints

\bibitem[\protect\citeauthoryear{{Carlberg}}{{Carlberg}}{2013}]{carlberg13}
{Carlberg} R.~G.,  2013, \apj, 775, 90

\bibitem[\protect\citeauthoryear{{Chernoff} \& {Weinberg}}{{Chernoff} \&
  {Weinberg}}{1990}]{chernoff90}
{Chernoff} D.~F.,  {Weinberg} M.~D.,  1990, \apj, 351, 121

\bibitem[\protect\citeauthoryear{{Choi}, {Weinberg} \& {Katz}}{{Choi}
  et~al.}{2007}]{choi07}
{Choi} J.-H.,  {Weinberg} M.~D.,    {Katz} N.,  2007, \mnras, 381, 987

\bibitem[\protect\citeauthoryear{{Eyre} \& {Binney}}{{Eyre} \&
  {Binney}}{2011}]{eyre11}
{Eyre} A.,  {Binney} J.,  2011, \mnras, 413, 1852

\bibitem[\protect\citeauthoryear{{Fardal}, {Babul}, {Geehan} \&
  {Guhathakurta}}{{Fardal} et~al.}{2006}]{fardal06}
{Fardal} M.~A.,  {Babul} A.,  {Geehan} J.~J.,    {Guhathakurta} P.,  2006,
  \mnras, 366, 1012

\bibitem[\protect\citeauthoryear{{Fardal}, {Weinberg}, {Babul}, {Irwin},
  {Guhathakurta}, {Gilbert}, {Ferguson}, {Ibata}, {Lewis}, {Tanvir} \&
  {Huxor}}{{Fardal} et~al.}{2013}]{fardal13}
{Fardal} M.~A.,  {Weinberg} M.~D.,  {Babul} A.,  {Irwin} M.~J.,  {Guhathakurta}
  P.,  {Gilbert} K.~M.,  {Ferguson} A.~M.~N.,  {Ibata} R.~A.,  {Lewis} G.~F.,
  {Tanvir} N.~R.,    {Huxor} A.~P.,  2013, \mnras, 434, 2779

\bibitem[\protect\citeauthoryear{{Gibbons}, {Belokurov} \& {Evans}}{{Gibbons}
  et~al.}{2014}]{gibbons14}
{Gibbons} S.~L.~J.,  {Belokurov} V.,    {Evans} N.~W.,  2014, ArXiv e-prints

\bibitem[\protect\citeauthoryear{{Gieles}, {Alexander}, {Lamers} \&
  {Baumgardt}}{{Gieles} et~al.}{2014}]{gieles14}
{Gieles} M.,  {Alexander} P.~E.~R.,  {Lamers} H.~J.~G.~L.~M.,    {Baumgardt}
  H.,  2014, \mnras, 437, 916

\bibitem[\protect\citeauthoryear{{Gnedin} \& {Ostriker}}{{Gnedin} \&
  {Ostriker}}{1997}]{gnedin97}
{Gnedin} O.~Y.,  {Ostriker} J.~P.,  1997, \apj, 474, 223

\bibitem[\protect\citeauthoryear{{Hayashi}, {Navarro}, {Taylor}, {Stadel} \&
  {Quinn}}{{Hayashi} et~al.}{2003}]{hayashi03}
{Hayashi} E.,  {Navarro} J.~F.,  {Taylor} J.~E.,  {Stadel} J.,    {Quinn} T.,
  2003, \apj, 584, 541

\bibitem[\protect\citeauthoryear{{Heggie}}{{Heggie}}{2001a}]{heggie01hill}
{Heggie} D.~C.,  2001a, in {Steves} B.~A.,  {Maciejewski} A.~J.,  eds, The
  Restless Universe {Escape in Hill's problem}.
pp 109--128

\bibitem[\protect\citeauthoryear{{Heggie}}{{Heggie}}{2001b}]{heggie01massloss}
{Heggie} D.~C.,  2001b, in {Deiters} S.,  {Fuchs} B.,  {Just} A.,  {Spurzem}
  R.,   {Wielen} R.,  eds, Dynamics of Star Clusters and the Milky Way Vol.~228
  of Astronomical Society of the Pacific Conference Series, {Mass Loss from
  Globular Clusters}.
p.~29

\bibitem[\protect\citeauthoryear{{Helmi} \& {White}}{{Helmi} \&
  {White}}{1999}]{helmi99}
{Helmi} A.,  {White} S.~D.~M.,  1999, \mnras, 307, 495

\bibitem[\protect\citeauthoryear{{Howley}, {Geha}, {Guhathakurta},
  {Montgomery}, {Laughlin} \& {Johnston}}{{Howley} et~al.}{2008}]{howley08}
{Howley} K.~M.,  {Geha} M.,  {Guhathakurta} P.,  {Montgomery} R.~M.,
  {Laughlin} G.,    {Johnston} K.~V.,  2008, \apj, 683, 722

\bibitem[\protect\citeauthoryear{{Hunter}}{{Hunter}}{2005}]{hunter05}
{Hunter} C.,  2005, Annals of the New York Academy of Sciences, 1045, 120

\bibitem[\protect\citeauthoryear{{Johnston}}{{Johnston}}{1998}]{johnston98}
{Johnston} K.~V.,  1998, \apj, 495, 297

\bibitem[\protect\citeauthoryear{{Just}, {Berczik}, {Petrov} \& {Ernst}}{{Just}
  et~al.}{2009}]{just09}
{Just} A.,  {Berczik} P.,  {Petrov} M.~I.,    {Ernst} A.,  2009, \mnras, 392,
  969

\bibitem[\protect\citeauthoryear{{King}}{{King}}{1966}]{king66}
{King} I.~R.,  1966, \aj, 71, 64

\bibitem[\protect\citeauthoryear{{K{\"u}pper}, {Kroupa}, {Baumgardt} \&
  {Heggie}}{{K{\"u}pper} et~al.}{2010}]{kuepper10}
{K{\"u}pper} A.~H.~W.,  {Kroupa} P.,  {Baumgardt} H.,    {Heggie} D.~C.,  2010,
  \mnras, 401, 105

\bibitem[\protect\citeauthoryear{{K{\"u}pper}, {Lane} \& {Heggie}}{{K{\"u}pper}
  et~al.}{2012}]{kuepper12}
{K{\"u}pper} A.~H.~W.,  {Lane} R.~R.,    {Heggie} D.~C.,  2012, \mnras, 420,
  2700

\bibitem[\protect\citeauthoryear{{K{\"u}pper}, {MacLeod} \&
  {Heggie}}{{K{\"u}pper} et~al.}{2008}]{kuepper08}
{K{\"u}pper} A.~H.~W.,  {MacLeod} A.,    {Heggie} D.~C.,  2008, \mnras, 387,
  1248

\bibitem[\protect\citeauthoryear{{Lamers}, {Baumgardt} \& {Gieles}}{{Lamers}
  et~al.}{2010}]{lamers10}
{Lamers} H.~J.~G.~L.~M.,  {Baumgardt} H.,    {Gieles} M.,  2010, \mnras, 409,
  305

\bibitem[\protect\citeauthoryear{{Murali} \& {Weinberg}}{{Murali} \&
  {Weinberg}}{1997}]{murali97}
{Murali} C.,  {Weinberg} M.~D.,  1997, \mnras, 291, 717

\bibitem[\protect\citeauthoryear{{Odenkirchen}, {Grebel}, {Dehnen}, {Rix},
  {Yanny}, {Newberg}, {Rockosi}, {Mart{\'{\i}}nez-Delgado}, {Brinkmann} \&
  {Pier}}{{Odenkirchen} et~al.}{2003}]{odenkirchen03}
{Odenkirchen} M.,  {Grebel} E.~K.,  {Dehnen} W.,  {Rix} H.,  {Yanny} B.,
  {Newberg} H.~J.,  {Rockosi} C.~M.,  {Mart{\'{\i}}nez-Delgado} D.,
  {Brinkmann} J.,    {Pier} J.~R.,  2003, \aj, 126, 2385

\bibitem[\protect\citeauthoryear{{Pe{\~n}arrubia}, {Navarro} \&
  {McConnachie}}{{Pe{\~n}arrubia} et~al.}{2008}]{jorge08}
{Pe{\~n}arrubia} J.,  {Navarro} J.~F.,    {McConnachie} A.~W.,  2008, \apj,
  673, 226

\bibitem[\protect\citeauthoryear{{Price-Whelan}, {Hogg}, {Johnston} \&
  {Hendel}}{{Price-Whelan} et~al.}{2014}]{pricewhelan14}
{Price-Whelan} A.~M.,  {Hogg} D.~W.,  {Johnston} K.~V.,    {Hendel} D.,  2014,
  ArXiv e-prints

\bibitem[\protect\citeauthoryear{{Sanders}}{{Sanders}}{2014}]{sanders14}
{Sanders} J.~L.,  2014, \mnras, 443, 423

\bibitem[\protect\citeauthoryear{{Sanders} \& {Binney}}{{Sanders} \&
  {Binney}}{2013a}]{sanders13a}
{Sanders} J.~L.,  {Binney} J.,  2013a, \mnras, 433, 1813

\bibitem[\protect\citeauthoryear{{Sanders} \& {Binney}}{{Sanders} \&
  {Binney}}{2013b}]{sanders13b}
{Sanders} J.~L.,  {Binney} J.,  2013b, \mnras, 433, 1826

\bibitem[\protect\citeauthoryear{{Sanderson}, {Helmi} \& {Hogg}}{{Sanderson}
  et~al.}{2014}]{sanderson14}
{Sanderson} R.~E.,  {Helmi} A.,    {Hogg} D.~W.,  2014, in {Feltzing} S.,
  {Zhao} G.,  {Walton} N.~A.,   {Whitelock} P.,  eds, IAU Symposium Vol.~298 of
  IAU Symposium, {Action-space clustering of tidal streams to map the Galactic
  potential}.
pp 207--212

\bibitem[\protect\citeauthoryear{{Siegal-Gaskins} \&
  {Valluri}}{{Siegal-Gaskins} \& {Valluri}}{2008}]{siegalgaskins08}
{Siegal-Gaskins} J.~M.,  {Valluri} M.,  2008, \apj, 681, 40

\bibitem[\protect\citeauthoryear{{Sohn}, {van der Marel}, {Carlin}, {Majewski},
  {Kallivayalil}, {Law}, {Anderson} \& {Siegel}}{{Sohn} et~al.}{2014}]{sohn14}
{Sohn} S.~T.,  {van der Marel} R.~P.,  {Carlin} J.~L.,  {Majewski} S.~R.,
  {Kallivayalil} N.,  {Law} D.~R.,  {Anderson} J.,    {Siegel} M.~H.,  2014,
  ArXiv e-prints

\bibitem[\protect\citeauthoryear{{Takahashi} \& {Portegies Zwart}}{{Takahashi}
  \& {Portegies Zwart}}{2000}]{takahashi00}
{Takahashi} K.,  {Portegies Zwart} S.~F.,  2000, \apj, 535, 759

\bibitem[\protect\citeauthoryear{{van den Bosch}, {Lewis}, {Lake} \&
  {Stadel}}{{van den Bosch} et~al.}{1999}]{vdbosch99}
{van den Bosch} F.~C.,  {Lewis} G.~F.,  {Lake} G.,    {Stadel} J.,  1999, \apj,
  515, 50

\bibitem[\protect\citeauthoryear{{Varghese}, {Ibata} \& {Lewis}}{{Varghese}
  et~al.}{2011}]{varghese11}
{Varghese} A.,  {Ibata} R.,    {Lewis} G.~F.,  2011, \mnras, 417, 198

\end{thebibliography}
\label{lastpage}
}
\end{document}